\documentclass[sigconf]{acmart}

\usepackage{xspace}
\usepackage{xcolor}
\usepackage{color}
\usepackage{colortbl}
\usepackage{makecell}
\usepackage{multirow}
\usepackage{geometry}

\usepackage{titlesec}

\usepackage{listings}
\usepackage[lined,boxed,vlined,ruled,linesnumbered]{algorithm2e}

\usepackage{subcaption}
\usepackage{graphicx}
\usepackage{balance}  
\usepackage{epsfig}
\usepackage{graphicx}
\usepackage{epstopdf}
\usepackage{latexsym}
\usepackage{enumitem}

\PassOptionsToPackage{noend}{algpseudocode}
\usepackage{algpseudocode}
\usepackage{pifont}
\usepackage{epsfig}
\usepackage{amssymb}

\usepackage{newtxmath}
\usepackage{amsmath}
\usepackage{amsfonts}
\usepackage{stmaryrd}
\usepackage{url}
\usepackage{multirow}
\usepackage{array}
\usepackage[normalem]{ulem}
\usepackage{color}
\usepackage{cleveref}
\usepackage{xspace}
\usepackage{mathtools}
\usepackage{soul}
\usepackage{listings}
\usepackage{enumitem}
\usepackage{xcolor}
\usepackage{tikz}
\newsavebox{\blackball}
\newsavebox{\greenball}

\usepackage[utf8]{inputenc}
\usepackage{enumitem}
\usepackage{amsmath}

\usepackage{bbding}
\usepackage{amssymb}
\usepackage{epstopdf}
\usepackage{epsfig,endnotes}
\usepackage{url}
\usepackage{array}
\usepackage{booktabs}
\usepackage{threeparttable}
\usepackage{tabulary}
\usepackage{CJKutf8}
\usepackage[most,breakable]{tcolorbox}
\usepackage{etoolbox}
\usepackage{fancyhdr}
\usepackage{lipsum}
\usepackage[fencedCode]{markdown}
\usepackage{tikz}
\definecolor{shadecolor}{rgb}{0.92,0.92,0.92}
\usepackage{hyperref}
\usepackage{balance}  
\usepackage{tikz}
\usepackage{lipsum} 
\usepackage{marginnote}

\usepackage{listings,upquote,soul,framed}
\colorlet{shadecolor}{gray!20}

\usepackage[lined,boxed,vlined,ruled,linesnumbered]{algorithm2e}

\usepackage{array}
\usepackage{longtable}

\newcommand{\zw}[1]{\textcolor{purple}{#1}}
\newcommand{\zxh}[1]{\textcolor{black}{#1}}

\newcommand{\blue}[1]{\textcolor{blue}{#1}}

\setcopyright{none}
\renewcommand\footnotetextcopyrightpermission[1]{} 
\newcommand\vldbdoi{XX.XX/XXX.XX}
\newcommand\vldbpages{XXX-XXX}
\newcommand\vldbvolume{14}
\newcommand\vldbissue{1}
\newcommand\vldbyear{2020}
\newcommand\vldbauthors{\authors}
\newcommand\vldbtitle{\shorttitle} 
\newcommand\vldbavailabilityurl{URL_TO_YOUR_ARTIFACTS}
\newcommand\vldbpagestyle{plain} 

\newcommand{\hi}[1]{\vspace{.25em} \noindent {\bf #1} }
\newcommand{\llm}{\textsc{LLM}\xspace}
\newcommand{\llms}{\textsc{LLMs}\xspace}
\newcommand{\bfit}[1]{\textbf{\textit{#1}}}
\newcommand{\oursys}{\texttt{OpenMLDB}\xspace}
\newcommand{\gpt}{\textsf{GPT-4o}\xspace}

\newcommand{\microbench}{\texttt{MicroBench}\xspace}
\newcommand{\realbenchone}{\texttt{RTP}\xspace}
\newcommand{\realbenchtwo}{\texttt{GLQ}\xspace}
\newcommand{\membench}{\texttt{TalkingData}\xspace}

\newcommand{\trino}{\texttt{Trino+Redis}\xspace}
\newcommand{\duckdb}{\texttt{DuckDB}\xspace}
\newcommand{\mysql}{\texttt{MySQL(in-mem)}\xspace}
\newcommand{\greenplum}{\texttt{GreenPlum}\xspace}
\newcommand{\flink}{\texttt{Flink}\xspace}

\usepackage{bm}

\theoremstyle{definition}
\newtheorem{definition}{Definition}[section]

\makeatletter
\newcommand{\removelatexerror}{\let\@latex@error\@gobble}
\makeatother

\begin{document}
\title{\oursys: A Real-Time Relational Data Feature Computation System for Online ML}

\author{Xuanhe Zhou}
\affiliation{\institution{{Shanghai Jiao Tong Univ.}}\country{}}
\email{{zhouxh@cs.sjtu.edu.cn}}

\author{Wei Zhou}
\affiliation{\institution{{Shanghai Jiao Tong Univ.}}\country{}}
\email{{weizhoudb@gmail.com}}

\author{Liguo Qi}
\affiliation{\institution{{4Paradigm Inc.}}\country{}}
\email{{qiliguo@4paradigm.com}}

\author{Hao Zhang}
\affiliation{\institution{{4Paradigm Inc.}}\country{}}
\email{{zhanghao@4paradigm.com}}

\author{Dihao Chen}
\affiliation{\institution{{SF Express Inc.}}\country{}}
\email{{chendh@sf-express.com}}

\author{Bingsheng He}
\affiliation{\institution{{National Univ. of Singapore}}\country{}}
\email{{hebs@comp.nus.edu.sg}}

\author{Mian Lu}
\affiliation{\institution{{4Paradigm Inc.}}\country{}}
\email{{lumian@4paradigm.com}}

\author{Guoliang Li}
\affiliation{\institution{{Tsinghua University}}\country{}}
\email{{liguoliang@tsinghua.edu.cn}}

\author{Fan Wu}
\affiliation{\institution{{Shanghai Jiao Tong Univ.}}\country{}}
\email{{fwu@cs.sjtu.edu.cn}}

\author{Yuqiang Chen}
\affiliation{\institution{{4Paradigm Inc.}}\country{}}
\email{{chenyuqiang@4paradigm.com}}

\pagestyle{plain}
\pagenumbering{arabic}

\begin{abstract}
Efficient and consistent feature computation is crucial for a wide range of online ML applications. Typically, feature computation is divided into two distinct phases, i.e., offline stage for model training and online stage for model serving. These phases often rely on execution engines with different interface languages and function implementations, causing significant inconsistencies. Moreover, many online ML features involve complex time-series computations (e.g., functions over varied-length table windows) that differ from standard streaming and analytical queries. Existing data processing systems (e.g., Spark, Flink, DuckDB) often incur multi-second latencies for these computations, making them unsuitable for real-time online ML applications that demand timely feature updates.

This paper presents \oursys, a feature computation system deployed in 4Paradigm's SageOne platform and over {100} real scenarios. Technically, \oursys first employs a \emph{unified query plan generator} for consistent computation results across the offline and online stages, significantly reducing feature deployment overhead. Second, \oursys provides an \emph{online execution engine} that resolves performance bottlenecks caused by long window computations (via pre-aggregation) and multi-table window unions (via data self-adjusting). It also provides a high-performance \emph{offline execution engine} with window parallel optimization and time-aware data skew resolving. Third, \oursys features a \emph{compact data format} and \emph{stream-focused indexing} to maximize memory usage and accelerate data access. Evaluations in testing and real workloads reveal significant performance improvements and resource savings compared to the baseline systems. The open community of \oursys now has over 150 contributors and gained 1.6k stars on GitHub\footnote{\blue{\url{http://github.com/4paradigm/OpenMLDB}}}.


\end{abstract}

\maketitle

\begin{figure}[!t]
  \centering
  \includegraphics[width=\linewidth]{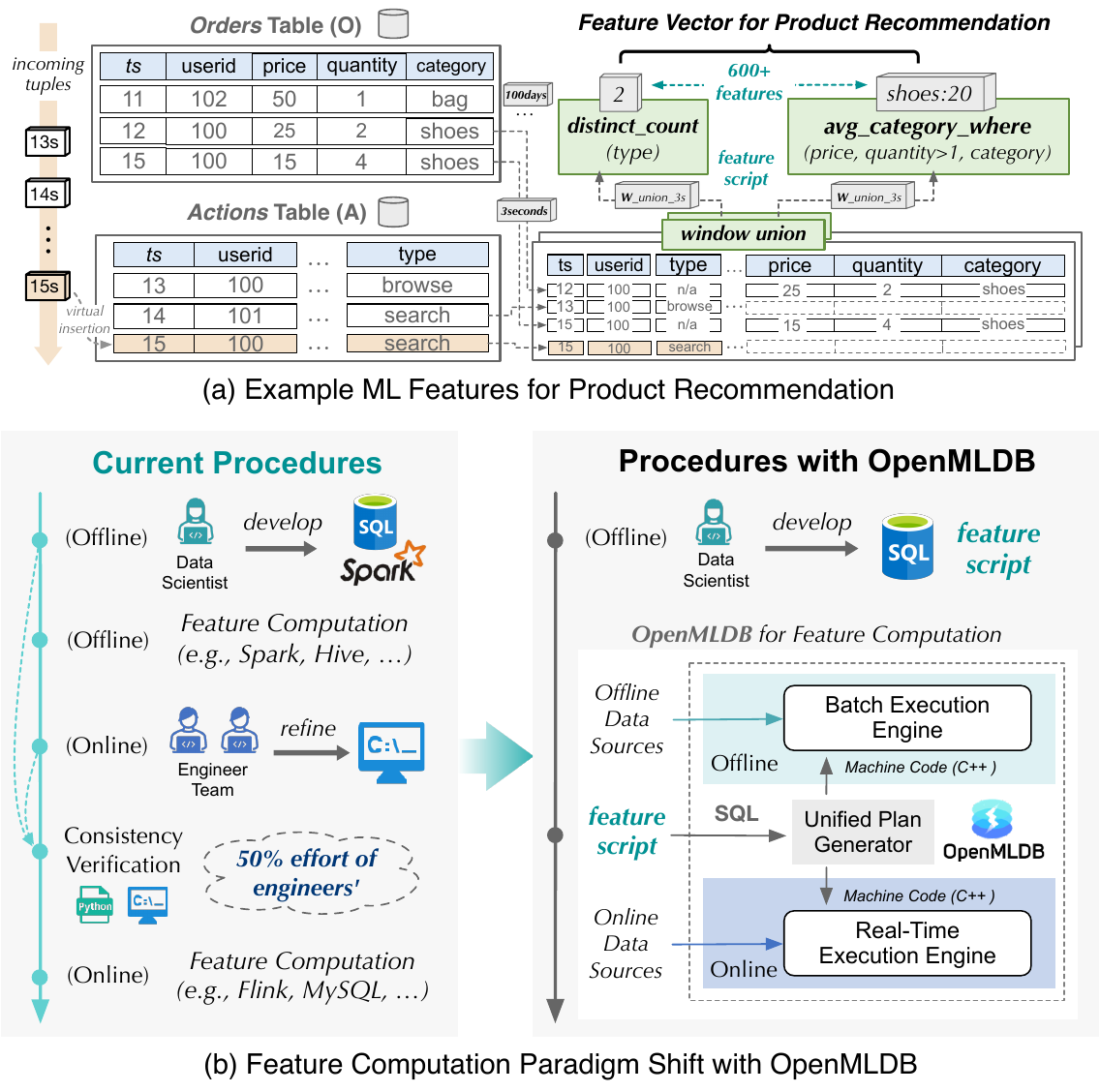}
    \vspace{-1em}
  \caption{Relational Data Feature Computation.}
  \label{fig:intro}
    \vspace{-1.25em}
\end{figure}






\section{Introduction}\label{sec:intro}

Relational data feature computation, which extracts ML features from raw data tables, is essential for many online applications leveraging machine learning models (e.g., credit card anti-fraud,  online advertising~\cite{ai2019,DBLP:journals/pvldb/SunL19,DBLP:journals/pvldb/WangCLL21}). However, this task is inherently complex. For instance, the product recommendation model at Vipshop requires over 600 distinct features~\cite{DBLP:journals/ijon/CaiLWY18}. Many of these features demand specialized ML functions, such as statistical functions (e.g., top-N frequency counts) and feature signatures (e.g., export format labeling for high-dimension features like product items~\cite{signature1}), as well as computation patterns (e.g., multi-table windows) distinct from those in standard streaming and analytical queries~\cite{febench,DBLP:conf/icde/ZhangZZLLZ23}.

{\it \bfit{Example.} To illustrate the complexity, consider the product recommendation scenario that generates personalized advertisements based on the user’s most recent actions. As shown in Figure~\ref{fig:intro}~(a), the feature computation process begins by computing window tables. For instance, we retrieve tuples from Actions and Orders that match the incoming tuple’s user ID and fall within a short time interval (e.g., three seconds). These matched tuples, together with the incoming tuple, form a window table ($w\_union\_{3s}$). Next, we compute various window features like $(i)$ counting the number of product types purchased by the user and $(ii)$ calculating the average price of products within specific categories, subject to conditions like ``purchased more than one''. To support recommendation at scale, feature computation must handle extremely high concurrency (e.g., more than {200 million} requests per minute) and still deliver up-to-date features within milliseconds.}

In current practice, a feature computation workflow has two separate stages. As shown in Figure~\ref{fig:intro}~(b), first, data scientists develop and compute offline features with batch processing engines (e.g., SparkSQL), iterating until the model’s training performance meets their objectives. Afterward, to deploy these offline feature scripts in online production, the engineering team must refactor and optimize them for online computation (with streaming engines like Flink). However, the two stages are handled by different teams and systems, which can cause inconsistencies. For instance, in a real-world de-fraud case at Varo~\cite{varo}, the online application defines the {\it ``account balance’’} feature as the current real-time balance, while offline data scientists define it as the previous day’s balance. Such discrepancies cause mismatches between offline and online results, making extensive and costly verification necessary before the final deployment. Based on our experience (e.g., at Akulaku~\cite{akulaku}), consistency verification can take several months or even one year.

\vspace{.5em}

Moreover, existing data processing systems, both online~\cite{apacheflink,sparkstream,mysql,mariadb,postgresql,hbase,cassandra,dynamodb,mongodb} and offline~\cite{zaharia2016apache,hive,mapreduce,doris,clickhouse,greenplum}, as well as specialized feature stores~\cite{feast,tecton,hopsworks,feathr}, struggle to meet the performance demands (see Section~\ref{subsec:relatedwork}).  For instance, streaming execution engines like Apache Flink \cite{apacheflink}, and Spark Streaming \cite{sparkstream} lack critical optimizations for feature computation, such as handling long window operations and aggregating data across multiple table windows, making it difficult to achieve sub-second online response time. Similarly, widely used storage engines (e.g., MySQL~\cite{mysql}, PostgreSQL~\cite{postgresql}, Cassandra~\cite{cassandra}) are not designed for rapid ingestion or efficient ordering of time-series data, both of which are essential for online feature storage. On the offline side, the execution engines (e.g., Spark~\cite{zaharia2016apache}, Hive~\cite{hive}, ClickHouse~\cite{clickhouse}) face issues including limited operation-level parallelism, data skew, and frequent RPC calls during distributed computation, which limit their throughput and scalability for offline feature computation. Lastly, existing feature stores like Feast~\cite{feast} primarily focus on fast data retrieval rather than real-time feature computation, forcing users to combine multiple tools for feature computation and thus introducing additional synchronization and operational overhead.

\vspace{.5em}

To address these problems, we propose \oursys, an industrial-level feature computation system. Specifically, \oursys first ensures \bfit{online-offline feature computation consistency} by providing a unified query plan generator that supports extended SQL syntax and compilation-level optimizations (e.g., cycle binding, compilation caching). Second, \oursys achieves \bfit{ultra-high feature computation performance} through a suite of optimization techniques for both online and offline computations. For online computation, it reuses pre-aggregation results to enhance long-window computations and employs a self-adjusting strategy to optimize multi-table window unions. For offline computation, it supports multi-window parallel optimization and mitigates data skew by repartitioning tuples based on utilized columns and data distribution. Finally, \oursys offers a compact in-memory data format and a stream-focused data structure, which pre-ranks data by key and timestamp to enable rapid online data access.

\vspace{1em}
We summarize our main contributions as follows.



$\bullet$  We present \oursys, an industry-level feature computation system recognized as a recommended feature product by Gartner~\cite{gartner-feature}, which has been deployed in 100+ real-world scenarios\footnote{See user feedback at {\blue{\url{https://github.com/4paradigm/OpenMLDB/discussions/707}}}.}. 


$\bullet$  We provide a unified query plan generator that streamlines feature script writing for offline and online stages and reduces feature deployment time from months to days or even less~\cite{developcost}.

$\bullet$ We propose an online real-time execution engine that resolves performance bottlenecks {in ML feature computations} such as long window aggregations and multi-table window unions.

$\bullet$ We propose an offline batch execution engine that tackles challenges such as low parallelism in multi-window aggregations and data-skewed partitionings in existing systems.

$\bullet$ We provide a compact in-memory data format for efficient memory usage and stream-oriented structures to optimize frequent data access during online feature computation.



$\bullet$ We thoroughly evaluated \oursys across various workloads. Our results show that \oursys delivers (1) 10x–20x higher online performance than Flink and DuckDB; (2) 6x faster offline performance than Spark and MPP databases like GreenPlum; and (3) 40\% lower memory usage compared to in-memory databases like Redis.

\vspace{-.5em}
\section{Background and Related Work}
\label{sec:motivation}

\subsection{Relational Data Feature Computation}
\label{subsec:define}

Given a set of data tables $\mathcal{R}$, the goal of relational data feature computation is to produce feature values, as specified in a feature script $Q$, that may include both discrete and continuous features. These feature computations often extend beyond standard relational operations (see Section~\ref{subsec:sql}) and present unique challenges. For example, they may require specialized windowed aggregations (e.g., top-frequency keys, conditional averages, time-series indicators), advanced string parsing, window unions, last-join operations, and direct output in ML-friendly formats. Based on distinct performance requirements, it can be further divided into online and offline stages.

\hi{Online Feature Computation} processes features for continuously arriving data tuples and provides up-to-date features for online requests. The computations draw on both recent data (often spanning windows of different time lengths) and static reference data (e.g., user profiles). Many commercial scenarios demand stringent real-time performance. For example, among the target clients of \oursys, \emph{Akulaku} (a fintech company) requires about 4-ms response time for risk control, and bank’s real-time anti-fraud checks must be completed under 20 ms~\cite{bank}. Similarly, companies such as \emph{UBiX} and \emph{Vipshop} need sub-100 ms latency for online advertising. 

\hi{Offline Feature Computation} processes features in batch mode against periodically updated tables (e.g., historical orders). The primary objective is to efficiently handle large batches of computation requests while maximizing system throughput. For instance, in consumer-facing scenarios like banks and \emph{Yum Holdings} (a restaurant company), daily active users reach tens of millions, and achieving higher throughput helps reduce machine costs. In a similar vein, when Akulaku adopted \oursys for offline risk control, they saved 20 servers (each with 128GB RAM), totaling around \emph{\$275,000}.



\vspace{-.5em}
\subsection{Existing Data Systems}
\label{subsec:relatedwork}

\hi{Online Execution Engines.} General-purpose stream processing systems~\cite{apacheflink,sparkstream} are insufficient for millisecond-level online ML services. Apache Flink~\cite{apacheflink} incurs significant overhead due to frequent context switching and lacks advanced optimizations for stateful aggregations. For instance, with sliding windows, Flink needs to select and discard the oldest data during each window operation. Since it lacks a robust state retention mechanism, Flink has to re-sort the data to identify the oldest entries for eviction, increasing the time complexity from O(1) to O(log n). Additionally, its Java-based implementation lacks specific low-level optimizations, resulting in high latency for individual requests. Similarly, Spark Streaming~\cite{sparkstream} relies on multi-node computation and requires frequent RPC calls to complete operations, making it challenging to achieve latencies in the tens-of-milliseconds range.

\begin{sloppypar}
\hi{Online Storage Engines.} 
Relational databases such as MySQL~\cite{mysql}, MariaDB~\cite{mariadb}, and PostgreSQL~\cite{postgresql}, even when configured for in-memory storage engines, often exhibit inefficiencies in standard indexing for time-based queries (e.g., rapid data insertion). They also lack native time-ordering capabilities essential for real-time analytics and sequential data processing. Distributed storage systems like HBase~\cite{hbase}, Cassandra~\cite{cassandra}, DynamoDB~\cite{dynamodb}, and MongoDB~\cite{mongodb} do not support standard ANSI SQL and lack optimizations for time-series data, which increases usage complexity for data scientists. While NewSQL systems (TiDB~\cite{tidb}, CockroachDB~\cite{cockroachdb}, OceanBase~\cite{oceanbase}) simplify usage via SQL support, they are also not optimized for time-series data. Meanwhile, time-series databases (e.g., InfluxDB~\cite{influxdata}, TDengine~\cite{tdengine},  OpenTSDB~\cite{opentsdb}) cannot fully manage data in memory space (primarily on disk), limiting their ability to support high-performance feature extraction.
\end{sloppypar}

\hi{Offline Execution Engines.} Existing data processing systems~\cite{zaharia2016apache,hive,mapreduce} lack specific optimizations for feature extraction. For instance, Spark's window aggregation does not support Whole-Stage Java Code Generation and struggles to parallelize many window operations without overlaps. Similarly, Apache Hive~\cite{hive}, built on top of MapReduce~\cite{mapreduce}, suffers from high job startup latency and disk-based operations, rendering it unsuitable for high-throughput iterative feature extraction. Other MPP systems, including Apache Doris~\cite{doris}, ClickHouse~\cite{clickhouse}, GreenPlum~\cite{greenplum}, Apache Druid~\cite{druid}, and StarRocks~\cite{starrocks}, are primarily designed for general analytical workloads rather than {time-series feature computations}. Furthermore, existing \emph{C++ execution engines} have their own limitations. Intel OAP~\cite{oap} does not extensively exploit advanced code generation techniques for complex transformations and aggregations. Databricks Photon~\cite{photon} and NVIDIA Spark RAPIDS~\cite{rapids} enable vectorized execution but lack specialized optimizations for unique feature computation patterns, such as efficient multi-window processing and custom aggregations.  Similarly, Alibaba EMR~\cite{alibabaemr} provides limited support for code generation in window functions, further restricting its applicability to feature extraction tasks.


\hi{Feature Stores.} Existing online feature stores such as Feast~\cite{feast} focus on rapid feature retrieval but lack mechanisms for on-the-fly, low-latency feature computation. Others, like Hopsworks~\cite{hopsworks}, Tecton~\cite{tecton}, and Feathr~\cite{feathr}, provide limited optimization for complex feature pipelines, especially those involving multiple data streams and windowed aggregations over relative time intervals. These limitations often force users to pre-process features using separate online storage systems and to re-synchronize whenever new features are introduced, resulting in operational overhead and delayed feature availability for online ML applications.

\section{Overview of \oursys}
\label{sec:overview}

\subsection{System Architecture}
\label{subsec:arch}

\begin{figure}[!t]
  \centering   \includegraphics[width=1\linewidth]{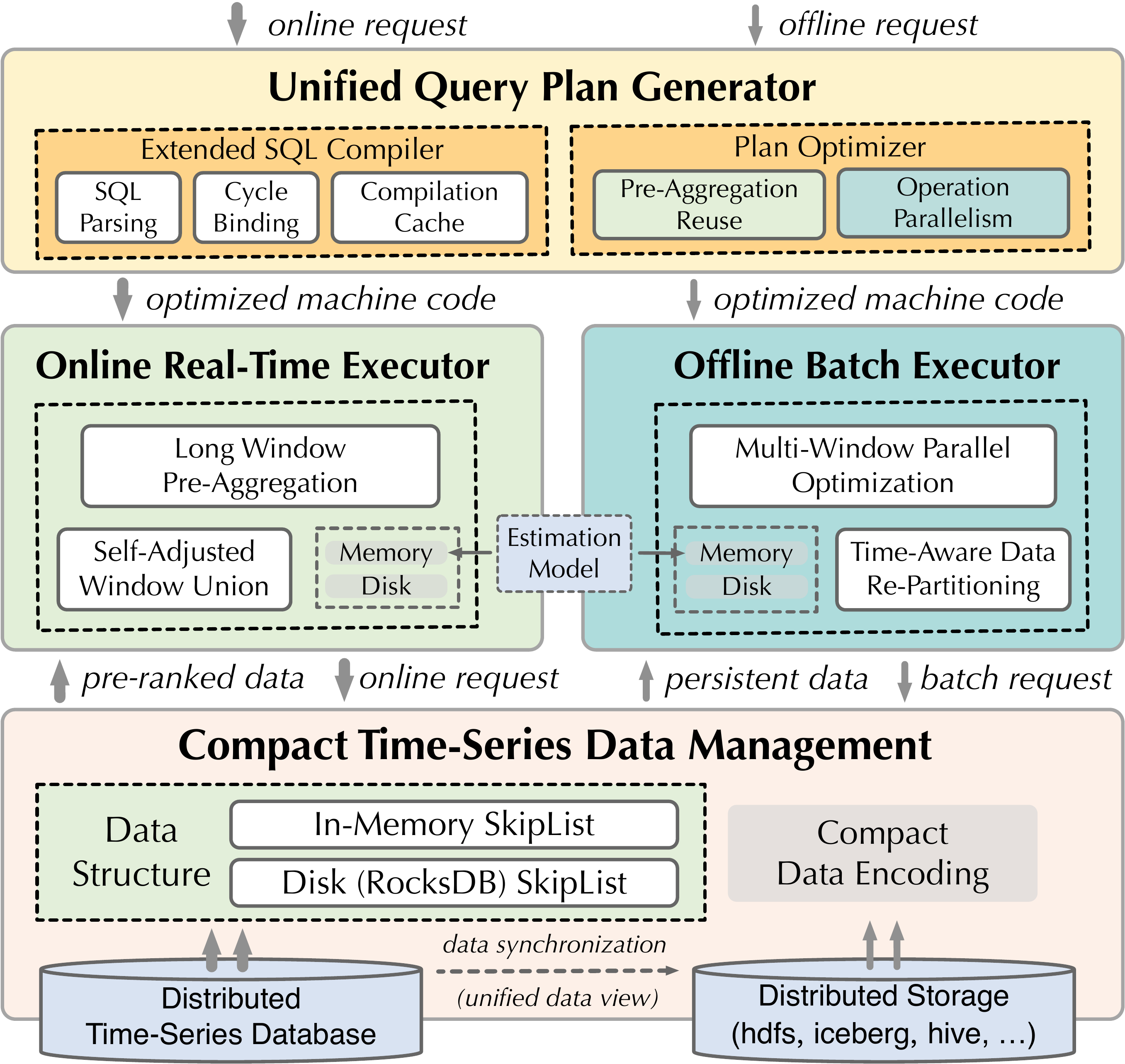}
  \caption{Architecture of \oursys.}
  \label{fig:arch}
  \vspace{-1.7em}
\end{figure}

As shown in Figure~\ref{fig:arch}, \oursys comprises a unified query plan generator, online real-time execution engine, offline batch execution engine, and compact time-series data and memory management. 

\hi{Unified Query Plan Generator.} To ensure consistency between offline and online feature computation, first our plan generator supports standard SQL for compatibility (e.g., MySQL queries) and SQL extensions to simplify the implementation of common features (see Section~\ref{sec:sql}). Second, our plan generator uses LLVM and JIT~\cite{DBLP:journals/corr/GregorS17,DBLP:journals/corr/abs-2008-05555,DBLP:conf/adc/MaYHCWJ23} to transform the SQL queries into efficient machine code, where we adopt optimizations like cycle binding and compilation caching. This machine code includes direct calls to the C++ library functions shared by the offline and online execution engines. 


\hi{Online Real-Time Execution.} We developed the online engine from scratch, which is several orders of magnitude faster than common stream processing and database systems. The underlying techniques include (1) \textit{pre-aggregation} for enhancing functions over windows with extremely long time intervals (e.g., for years) or hotspot data and (2) \textit{dynamic data adjusting} for efficient stream unions over multiple tables (e.g., with out-of-order stream data~\cite{10184828}). Note that we also adopt Zookeeper (a distributed framework~\cite{DBLP:conf/usenix/HuntKJR10}) to ensure high system availability and fault tolerance.


\hi{Offline Batch Execution.} Compared to engines like Spark, the offline engine in \oursys (1) improves the resource usage by parallelizing the execution of window operations over the same tables and (2) resolves the data skew problem by dynamically re-assigning window data based on the key columns and data distribution.


\hi{Compact Time-Series Data Management.} Compared with existing systems like Spark and Redis, \oursys supports a more compact data encoding format, which saves unnecessary memory consumption and enhances performance by allowing more data to be processed within limited memory space. Besides, \oursys proposes an effective in-memory data structure specifically for accelerating online feature operations like retrieving one latest tuple matched with the incoming one. 

\begin{figure}[!t]
  \centering   \includegraphics[width=1\linewidth]{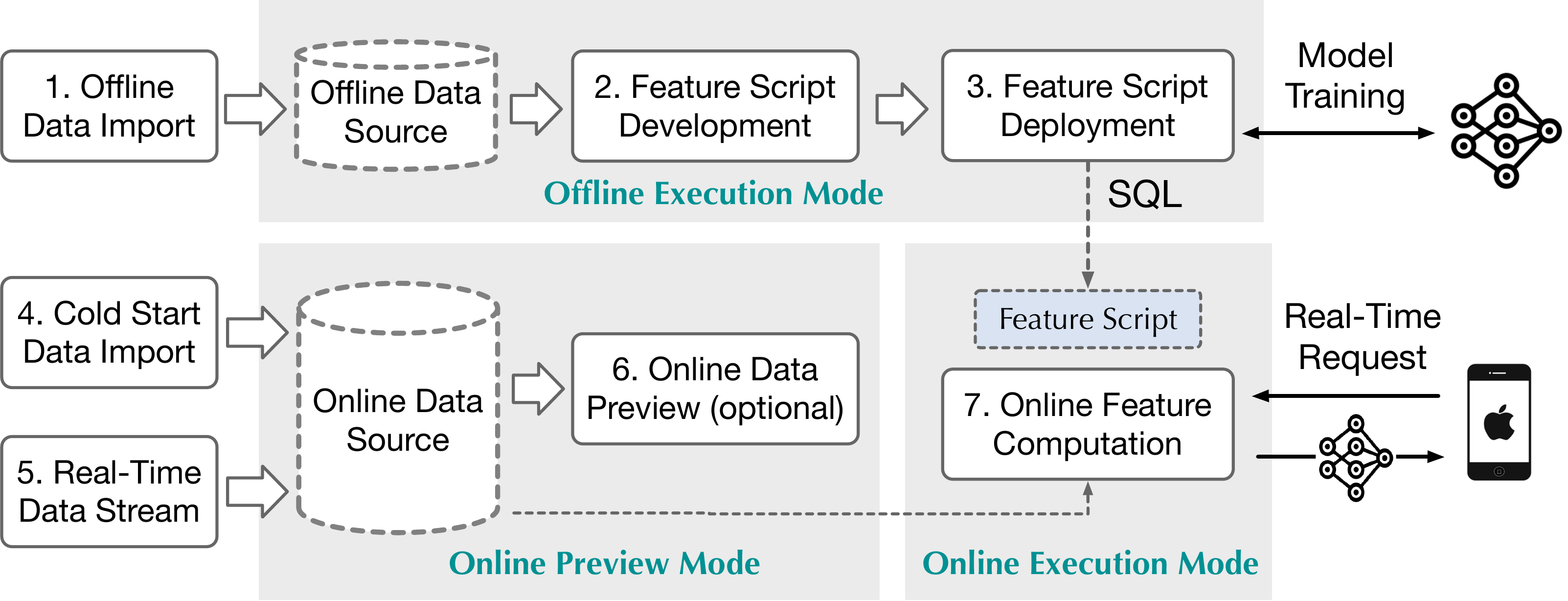}
  \vspace{-1.25em}
  \caption{Workflow of Different Execution Modes.}
  \label{fig:workflow}
  \vspace{-1.75em}
\end{figure}

Besides, to optimize memory usage, \oursys offers an empirical analysis model for memory estimation, which guides memory configuration (e.g., determining the machine number and settings) and table allocation across storage engines. Besides, \oursys provides a memory isolation and alerting mechanism to ensure availability when exceeding the memory limit.






\subsection{Execution Modes}
\label{subsec:workflow}

As illustrated in Figure~\ref{fig:workflow}, \oursys supports multiple execution modes for offline and online feature computation. These modes share the same SQL syntax but impose distinct execution constraints. \bfit{(1) Offline Execution Mode:} In this mode, historical data is synchronized with the offline engine, where data scientists develop feature scripts and execute large-scale computations. Certain SQL commands (e.g., \textsf{LOAD DATA}) are executed asynchronously as background tasks. \bfit{(2) Online Preview Mode:} This mode enables testing newly developed feature scripts on a limited set of online data, minimizing any negative impact on online services. To ensure efficiency, it retrieves results from a data cache and constrains query complexity (e.g., limiting the number of key columns). \bfit{(3) Online Request Mode:} In this mode, each incoming request tuple is virtually inserted into the online data table. The online engine then applies the previously deployed offline script and returns a single feature result for each tuple.




\section{Unified Query Plan Generator}
\label{sec:sql}


\subsection{\oursys SQL}  
\label{subsec:sql}  

SQL is one of the most popular data processing languages. Compared to other declarative languages like Python, most relational databases like MySQL and data processing systems like Spark support SQL, making it a common choice for data processing and feature extraction in ML workflows. Additionally, the inherent constraints of SQL (e.g., execution plans without arbitrary loops) enable more targeted optimizations~\cite{sparksql,calcite}.

\begin{table}[!t]
\small
\centering
\caption{Example \oursys Operations (w/o in Standard SQL).}
\label{table:openmldb_syntax}
\vspace{-.5em}
\begin{tabular}{|p{1.2cm}|p{6.4cm}|}
\hline
\textbf{Operation} & \textbf{\oursys} \\ \hline
\textbf{Window Function} &
\begin{tabular}[t]{@{}l@{}}
    \bfit{topN\_frequency}(column, top\_n) \\
    \bfit{avg\_cate\_where}(column, condition, category) \\
    \bfit{drawdown}(column) \\
    \bfit{ew\_avg}(column, smoothing factor) \\
    \bfit{split\_by\_key}(input\_string, delimeter, kv\_delimeter) \\
    \bfit{multiclass\_label}(column)
\end{tabular} \\
\hline
\textbf{Window Union} &
\begin{tabular}[t]{@{}l@{}}
    window\_clause: \\
    \hspace{5mm} \textbf{WINDOW} window\_expression [, ...] \\
    window\_expression: \\
    \hspace{5mm} named\_win \textbf{AS} \{ named\_win | ( window\_spec ) \} \\
    window\_spec: \\
    \hspace{5mm} [ \textbf{UNION} ( from\_item [, ...] ) ] \\
    \hspace{5mm} \textbf{PARTITION BY} expression [ \textbf{ORDER BY} ... ] \\
    \hspace{5mm} window\_frame\_clause [ window\_attr [, ...] ]
\end{tabular} \\
\hline
\textbf{Stream Join} &
\begin{tabular}[t]{@{}l@{}}
    join\_operation: \\
    \hspace{5mm} condition\_join\_operation \\
    condition\_join\_operation: \\
    \hspace{5mm} from\_item \textbf{LAST JOIN} [ \textbf{ORDER BY}... ] from\_item \\
    \hspace{5mm} join\_condition \\
    join\_condition: \\
    \hspace{5mm} \bfit{ON} bool\_expression
\end{tabular} \\
\hline
\end{tabular}
\vspace{-1.5em}
\end{table}

However, standard ANSI SQL~\cite{sqlstandard} has functional limitations. First, it lacks many ML feature functions {(e.g., max decline percentage from a historical peak)}. Second, standard {relational} operations are inefficient in handling ML data (e.g., with ultra-high-dimension sparse features). Thus, after applying standard SQLs, users generally need to further process the output features, such as writing Python Pandas to transform into formats (e.g., LibSVM format~~\cite{libsvm}) compatible with the ML frameworks (e.g., Scikit-learn, TensorFlow). To address these problems, \oursys extends standard SQL to include specialized feature computation operations\footnote{See \oursys SQL syntax with over 150 built-in functions for various ML feature types at \blue{\url{https://openmldb.ai/docs/en/main/openmldb\_sql/sql\_difference.html}}.}. Practical experience demonstrates that \oursys SQL is capable of supporting feature computation in various online ML tasks~\cite{akulaku}.  Here we showcase some operations in \oursys SQL (see Table~\ref{table:openmldb_syntax}).




\hi{Window Function.} Compared with basic aggregations in standard SQL, \oursys integrates a rich set of window functions for ML models. The following lists five typical categories:

\emph{(1) Frequency-Based Aggregations:} The \emph{topN\_frequency} function returns the top N keys sorted by their occurrence frequency. This requires complex implementation in standard SQL, involving multiple nested queries with group and ordering operations.

\begin{sloppypar}
\emph{(2) Conditional Aggregations:} Similar to streaming systems that prohibit full table scans (slowing down overall efficiency), \oursys extends some functions to support conditional filtering. For instance, the \emph{avg\_category\_where(value, condition, category)} function computes the average of values (in string format like {``shoes:20''}) that meet a specified \emph{condition}, grouped by the \emph{category} key. Achieving similar functionality in standard SQL would require more complex constructs, involving \textsf{CASE}, \textsf{WHERE}, and \textsf{ORDER BY} clauses.
\end{sloppypar}

\emph{(3) Time-Series Aggregations:} The \emph{drawdown} function computes the maximum decline percentage from a historical peak to a subsequent trough, which is used in scenarios like quantitative trading to measure maximum loss. This function is not provided by standard SQL. The \emph{ew\_avg} function calculates exponentially weighted value average, which requires pre-ordered data values (see Section~\ref{subsec:skiplist}).

\emph{(4) String Parsing:} The \emph{split\_by\_key} function splits a string by a delimiter, treats each segment as a key-value pair, and adds each key to an output list, which standard SQL does not natively support.


\emph{(5) Feature Signatures:} We support signature functions to mark column types by their usage in ML: $(i)$ Label Columns are directly retained without modification. $(ii)$ Discrete Columns are processed using a hash algorithm to generate high-dimensional feature values~\cite{columnhash}. $(iii)$
Continuous Columns retain their original values to produce one-dimensional dense features. In this way, we can directly output the features in {LIBSVM or TFRecords} format (for ML usage), avoiding to export raw table data with {ultra high} dimensions, such as millions of dimensions for product items. 

Note these functions can be accelerated for lengthy table windows using the pre-aggregation technique (see Section~\ref{subesc:pre-agg}).


\hi{Window Union.} The \textsf{WINDOW UNION} clause aggregates
data over a time window that includes data tuples from two or more tables (optimized via the self-adjusting technique in Section~\ref{subsec:interval-join}). This simplifies the query by avoiding complex \textsf{UNION ALL} operations and additional labels to distinguish which tables the tuples origin.

\hi{Stream Join.}  The \textsf{LAST JOIN} operation enables the matching of one most recent data tuples with the incoming data tuple (accelerated via the in-memory data structure in Section~\ref{subsec:skiplist}). This eliminates additional rank and filter operations in standard SQL. 

With these operations, we can efficiently implement the features in Figure~\ref{fig:intro}. In the SQL snippet, we first define time windows like $(i)$ \emph{w\_union\_{3s}} that partitions tuples by user ids (\emph{PARTITION BY userid}), orders them by timestamp (\emph{ORDER BY ts}) within a three-second interval before the incoming tuple (\emph{ROWS BETWEEN 3s $\cdots$}) and $(ii)$ long window \emph{w\_action\_{100d}} for the interval of 100 days. Upon \emph{w\_union\_{3s}}, we compute features like $(i)$ the number of distinct product types (\emph{distinct\_count(action.type)}) and $(ii)$ using the  \emph{avg\_category\_where} function to conditionally compute average product prices.

\lstdefinelanguage{SQL}{
  keywords={SELECT,FROM,WHERE,GROUP,BY,ORDER,OVER,WINDOW,PARTITION,UNION,AS,ROWS,BETWEEN,PRECEDING,AND,CURRENT,ROW},
  keywordstyle=\color[RGB]{83,128,53}\bfseries,
  sensitive=false,
  comment=[l]{--},
  commentstyle=\color{teal}\ttfamily,
  stringstyle=\color{orange}\ttfamily,
  morestring=[b]'
}

\lstset{
  language=SQL,
  escapechar=@,  
  basicstyle=\ttfamily,
  columns=fullflexible,
  showstringspaces=false,
  breaklines=true,
  frame=none   
}

\begin{lstlisting}
SELECT action.*,
    @\textbf{distinct\_count}@(action.type) AS product_count,
    @\textbf{avg\_category\_where}@(price, quantity > 1, category) OVER w_union_3s AS product_prices, ...
FROM actions WINDOW 
    w_union_3s AS (
      UNION orders PARTITION BY userid
      ORDER BY ts
      ROWS BETWEEN 3s PRECEDING AND CURRENT ROW),
    w_action_100d AS ( ... ), ...;
\end{lstlisting}

\subsection{SQL Compilation}
\label{subsec:compile}

We compile SQL into efficient C++ machine code by applying optimizations in three main aspects. \bfit{(1) Parsing Optimization:} \oursys first identifies the identical column references and similar window definitions within the deployed SQL. Instead of generating separate code segments for each instance, it merges them into a unified code block. For instance, when multiple windows share the same computation template (e.g., in the form like \textsf{``PARTITION BY key1 ORDER BY ts ROWS ...''}), they can be merged into a single window definition, preventing redundant window construction and reducing repetitive code generation. Besides, \oursys applies index optimizations to critical information (e.g., data provider) in \textsf{WINDOW} and \textsf{LAST JOIN} operations (see Section~\ref{subsec:skiplist}), so as to accelerate the execution process. \bfit{(2) Cyclic Binding:} When multiple aggregate functions appear in the same window, \oursys applies a cyclic binding strategy. First, it compiles the minimal intermediate states required by simpler aggregates like \emph{sum}, \emph{min}, and \emph{max}. These intermediate results are then reused to compute more complex aggregates such as \emph{avg}. \bfit{(3) Compilation Cache:} \oursys employs a caching mechanism that stores previously compiled code segments, allowing subsequent deployments of similar SQLs (feature scripts) to bypass the full compilation process. When a new SQL shares code patterns with an existing cached segment, \oursys retrieves the relevant compiled portion and directly applies it. This reuse of compiled artifacts accelerates computation by reducing the time spent on repeated code generation and optimization.

\section{Online Feature Computation}
\label{sec:online}

\begin{sloppypar}
Beyond compiling optimization, \oursys's online execution engine offers specific optimization techniques for time-consuming operations in \oursys SQL, including pre-aggregation for long window computation in Section~\ref{subesc:pre-agg} and self-adjusting computation for multiple window table unions in Section~\ref{subsec:interval-join}.
\end{sloppypar}

\subsection{Long Window Pre-Aggregation}

\label{subesc:pre-agg}

We first explain how to resolve the performance bottleneck caused by window functions involving extremely large data volumes. Such cases arise  $(i)$ when the time intervals are extremely long (e.g., spanning multiple years), which is uncommon in typical streaming scenarios, or $(ii)$ when containing hotspot data. We refer to them as \emph{long-window features}. On the one hand, as data volumes grow, the computation time for these long-window features increases proportionally, making it difficult to achieve real-time performance. On the other hand, processing overlapped long-window features will cause significant resource wastes.

\begin{figure}[!t]
  \centering
  \includegraphics[width=.95\linewidth]{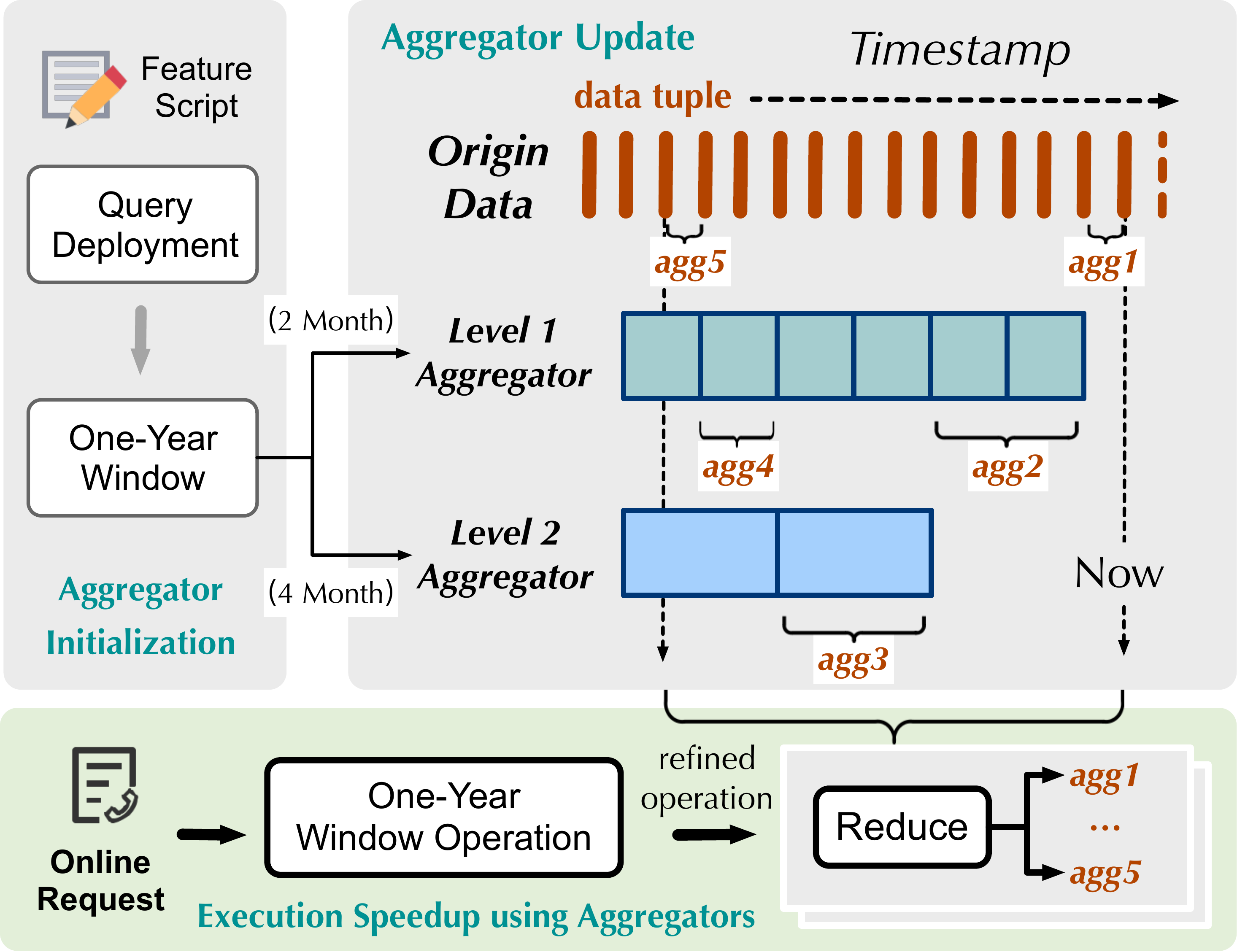}
  \caption{Long Window Pre-Aggregation.}
  \label{fig:pre-aggregate}
\end{figure}

\begin{sloppypar}
To address these problems, we propose a long-window pre-aggregation technique. As illustrated in Figure~\ref{fig:pre-aggregate}, we partially compute aggregation results during data insertion. Then, for real-time requests involving \emph{long-window features}, we simply compute by merging these pre-aggregated values, avoiding to process all raw tuples in the windows. Specifically, it includes three main steps.
\end{sloppypar}

\hi{Aggregator Initialization.} We first initialize multi-level aggregators based on the long-window features in the deployed scripts. Each level corresponds to a specific time granularity and builds upon the pre-aggregated results of finer-grained levels. We determine the optimal \emph{aggregator hierarchy} by considering factors such as query frequency, time window sizes, data distribution, and maintenance overhead. Meanwhile, we utilize data structures like segment trees~\cite{segmenttree} to efficiently manage historical computation results of these aggregators, based on which we compute statistics such as query frequencies for \emph{aggregator hierarchy} enhancement.

\hi{Query Refinement.} For an online request, the query optimizer identifies long-window computations and replaces expensive raw data scans with efficient lookups in the \emph{aggregator hierarchy}. As illustrated in Figure~\ref{fig:pre-aggregate}, the long-window operation is transformed into the merging of $(i)$ pre-aggregated features \emph{agg2} and \emph{agg4} from the daily aggregator (first level), $(ii)$ the pre-aggregated feature \emph{agg3} from the monthly aggregator (second level), and $(iii)$ real-time computed features from raw data \emph{agg1} and \emph{agg5}. 
Based on the evolving query patterns, \oursys can adaptively adjust the hierarchy by adding or removing aggregation levels, avoiding unnecessary computational or storage costs. For instance, to compute a feature over the past year, we can adopt aggregators at daily and monthly levels if hourly aggregators are seldom queried. 


\hi{Aggregator Update.} To maintain up-to-date aggregators without negatively impacting online data insertion, we employ an asynchronous updating strategy. First, aggregator updates are designed by assuming the ``binlog\_offset'' increases monotonically. To ensure this, all updates are protected within the replicator lock, which prevents other operations, such as concurrent \textsf{Put} requests, from inserting conflicting updates in the middle of the sequence. By coordinating updates through the binlog, we decouple pre-aggregation computations from the main data insertion pipeline. 
Besides, to support failure recovery during aggregation updates with binlog, we encapsulate the aggregator update logic within update\_aggr closure, ensuring atomic updates within the binlog sequence. Updates are appended to the binlog (``replicator->AppendEntry(entry, \&closure)''), which not only appends the updates but also triggers asynchronous execution.

\subsection{Self-Adjusted Window Union}
\label{subsec:interval-join}

\begin{sloppypar}
Next, we explain how to optimize multi-table window unions (\textsf{Window Union}), which are one of the most complex operations in online feature computation. \textsf{Window Union} aims to match tuples from multiple stream tables over a shared time window and partitioned by common keys. 
For example, as shown in Figure~\ref{fig:intro}, a 3-second \emph{Window Union} ($w\_union\_{3s}$) returns tuples arriving within the last three seconds to the current tuple. A standard approach, used by Flink~\cite{apacheflink}, employs a static key-based distribution strategy, where data tuples are routed to threads solely based on their hashed key columns. This rigid approach struggles to handle workload shifts effectively. To address this shortcoming, \oursys incorporates a self-adjusting technique that dynamically refines task allocation and applies incremental computations, ensuring low response time under various workloads. It includes two main steps.  \bfit{(1) On-the-Fly Load Balancing.} Our prior work~\cite{10184828} found that if the set of distinct keys is limited, static key-based distribution often encounters severe load imbalances. To alleviate this problem, we introduce a dynamic scheduler that periodically adjusts the mapping from keys to worker threads. By gathering runtime metrics on processing overhead, We estimate each worker thread’s load and reassign specific keys at runtime, keeping the workload even across all threads. Furthermore, multiple workers can collaborate on the same key subset if needed, improving flexibility and overall throughput. \bfit{(2) Incremental Computation.}  Large windows commonly create overlapping intervals of data, making computations repetitive and resource-intensive. To avoid recalculating results from scratch, we utilize a \emph{Subtract-and-Evict} approach~\cite{10.1145/3093742.3093925}. As an outdated tuple leaves the window, we subtract it from the running aggregator. Then, we incorporate any newly arriving tuples incrementally. 
\end{sloppypar}

\section{Offline Feature Computation}
\label{sec:offline}




\subsection{Multi-Window Parallel Optimization}
\label{subsec:parallel-optimize}

The first optimization in our offline engine focuses on \emph{multi-window parallel optimization}, which enhances the performance of queries involving different window functions over the same table. Consider a query that involves two window functions without dependencies: 
(1) \bfit{Window 1 (w1)} partitions rows by ``name'' and orders them by ``age''. (2) \bfit{Window 2 (w2)} partitions rows by ``age'' and orders them by ``age''. In traditional data processing systems, such queries are serially computed, even when the windows have no dependency relations. This sequential processing is inefficient and presents significant optimization opportunities. 
\oursys parallelizes these window operations by introducing two additional node types into the execution plan: ``Concat Join'' and ``Simple Project''. The ``Concat Join'' node concatenates features computed for the same data rows across different windows, marking the end of a parallel optimization segment. The ``Simple Project'' node represents the target columns without additional computations, marking the start of a parallel optimization segment.

Realizing parallel window computation is challenging because the order of data tuples can differ in windows using different partition keys (e.g., ``name'' for {\it w1} and ``age'' for {\it w2}). This discrepancy makes it difficult to align and concatenate the results from parallelized windows based on their natural order. To overcome this, we introduce an \emph{index column} to the input data, assigning a unique identifier to each data tuple. This index serves as a consistent join key across all window operations, ensuring correct alignment during concatenation regardless of the partitioning scheme.

Specifically, given the logical plan of a query, we first identify operation nodes that require parallel optimization. We perform a depth-first traversal from the root node to locate all ``Concat Join'' nodes inserted by the SQL parser to mark points needing parallel optimization. For each ``Concat Join'' node, we mark its child nodes $\{w_i\}$ (the original window operations) to include the index column in their outputs. We also find the nearest common ancestor node of $\{w_i\}$, which is a ``Simple Project'' node. At this node, we execute a `Column Add` action to insert the index column into the data stream table of these windows.

Next, with these labeled nodes, we execute the plan in a bottom-up manner. That is, each window function $w_i$ runs independently and outputs its computed features along with the index column. After all window operations have been completed, the ``Concat Join'' node concatenates their results using a ``Last Join'' operation, with the index column as the join key (optimized via indexes). Finally, the ``Concat Join'' node outputs the concatenated results and removes the index column to restore the original schema.

\subsection{Time-Aware Data Skew Resolving}
\label{subsec:repartition}

The second optimization aims to resolve the data skew problem in distributed feature computation. In the distributed framework of \oursys, when window computation encounters partition skew, and an imbalance is observed with excessive data in a particular partition, traditional data skew optimization methods like adding a random prefix to keys before repartitioning (known as ``salting'')~\cite{DBLP:conf/saci/HughesA21,salting} cannot be applied. Randomly prefixing keys can assign data tuples with the same group key to different partitions, leading to out-of-order processing and even inaccurate results.

To tackle this challenge, \oursys introduces a data-aware parallel computation strategy that dynamically adjusts data partitions based on workload and data distribution with the following steps. 

\bfit{Determine Partition Boundaries.} Given the historical data, we first evaluate the overall data to understand the distribution of the partition keys (e.g., \textsf{Gender}). We calculate statistical metrics to determine appropriate partition boundaries: \emph{(1) Quantile} determines how to split the data into $n$ equal parts. For example, setting \textsf{Quantile} to 4 aims to divide the data into four partitions; \emph{(2) Percentile (\textsf{PERCENTILE\_i}):} Based on the \textsf{Timestamp} column (e.g., used in the \texttt{ORDER BY} clause), we calculate boundary values that define the ranges for each partition. Data where \textsf{Timestamp} falls within $(\textsf{PERCENTILE\_i}, \textsf{PERCENTILE\_{i+1}}]$ belongs to the $i$-th partition. Note we adopt HyperLogLog~\cite{hyperloglog} algorithm to approximate the percentile distribution to avoid full data scan.


\bfit{Assign Repartitioning Identifiers.} Based on the distribution analysis, we tag each data tuple with: \emph{(1) PART\_ID:} A new partition identifier that, along with the original partition key (e.g., \textsf{Gender}), determines the new partitions after repartitioning. \emph{(2) EXPANDED\_ROW:} A boolean flag indicating whether the row is an original data row (\textsf{false}) or an expanded window row (\textsf{true}). In this way, we can control the repartitioning process without altering the original keys, thus preserving the continuity required for accurate window computations.

\bfit{Augment Window Data for Each Partition.} To ensure that window computations produce correct results after repartitioning (origin tuples in the same window may be of different partitions), we expand each partition's data by including additional data tuples from preceding partitions that are necessary for computing the window functions. The process involves: \emph{(1) Identify Required Data}: For each partition (except the first), we identify data tuples from earlier partitions that fall within the window frame needed for computations in the current partition. \emph{(2) Mark Expanded Data}: We set the \textsf{EXPANDED\_ROW} flag to \textsf{true} for these additional data tuples to distinguish them from the original data. \emph{(3) Union of Data}: We merge the expanded data tuples with the original partition data, ensuring each partition has all the data for accurate results.


\bfit{Redistribute Data by New Identifiers.} With the data tagged and expanded, we repartition the dataset based on both the original partition key and the new \textsf{PART\_ID}. This {\it increases the level of parallelism} by distributing the workload more evenly across multiple partitions. For example, if we split the data into four partitions, we can achieve a parallelism level of up to 8 when considering both \textsf{Gender} values and \textsf{PART\_ID}s.

\bfit{Compute on Repartitioned Data.} Finally, we perform the window computations on the repartitioned data. We compute only on data tuples where \textsf{EXPANDED\_ROW} is \textsf{false}, and the expanded data tuples (\textsf{EXPANDED\_ROW} is \textsf{true}) are utilized to provide the necessary context (e.g., identifying partition boundaries) without including them in the output.

\begin{figure}[!t]
  \centering
  \includegraphics[width=1\linewidth]{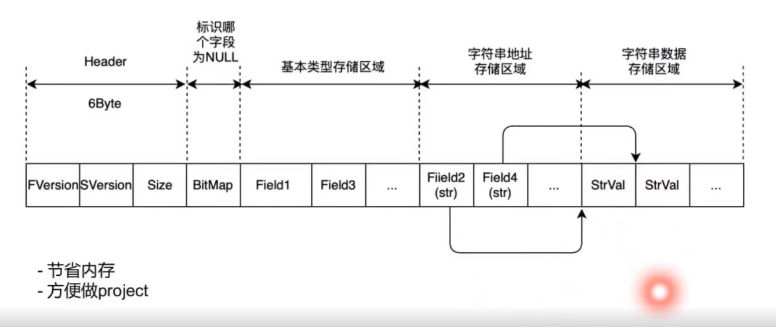}
  \vspace{-1.5em}
  \caption{In-Memory Data Encoding Format.}
  \label{fig:encoding}
  \vspace{-1.25em}
\end{figure}

\section{Compact Time-Series Data Management}
\label{sec:data}

\subsection{In-Memory Data Encoding}

\oursys uses a compact row encoding format that efficiently stores various data types while reducing memory usage. The data format combines fixed-size and variable-size data structures (see Figure~\ref{fig:encoding}). Each row consists of four parts. \bfit{(1) Header (6 bytes):} The row begins with a header that encodes essential metadata, including field version, schema version, and total row size. With fewer than 64 versions, each version requires only one byte and a 32-bit value stores the row’s size. \bfit{(2) BitMap:} After the header, the BitMap indicates which columns contain \textsf{NULL} values. Allocated in byte units, it marks \textsf{NULL} fields by setting specific bits, thus avoiding the direct storage of \textsf{NULL} values. Its size depends on the number of columns and the presence of \textsf{NULL}s. \bfit{(3) Basic Type Data Fields:} Next come the basic data types (e.g., integers, floats, timestamps), stored contiguously in memory at variable lengths. For integer data, \oursys uses a more compact offset calculation approach, speeding up offset determination and reducing storage costs. \bfit{(4) Variable-length Fields:} Finally, variable-length fields (e.g., strings) are stored by their offsets rather than embedding actual values. The string length is the difference between its current and previous offsets, and \oursys avoids unnecessary fixed-size (32-bit) length fields. Instead, it allocates precisely as much memory as the string requires, further optimizing space utilization.

Moreover, we utilize JNI to invoke Java's UnsafeRow data format~\cite{unsaferow}, which can directly read data Rows in C, access pointers, retrieve the NullMap and each column, and obtain values through pointer offsets. In this way, we reduce the need for row encoding and decoding transformations.

\hi{Memory Saving Example.} Consider an example row containing 20 integers, 20 floats, 20 strings (each with 1 byte), and 5 timestamps. In Spark, each row uses a \textsf{NULL} bit set (16 bytes), with each integer and float occupying 8 bytes, each string occupying 9 bytes (8 for data plus 1 for metadata), and each timestamp occupying 8 bytes. The total Spark row size is calculated as $556$ bytes. In contrast, \oursys’s row format includes a header (6 bytes), a null bit set (9 bytes), with each integer and float occupying 4 bytes, each string occupying 2 bytes (1 for data plus 1 for metadata), and each timestamp occupying 8 bytes. The total \oursys row size is \(255\) bytes, with over 54\% memory saving compared to Spark.




\subsection{In-Memory Data Structure}
\label{subsec:skiplist}


For online feature computation, traditional data structures like hash tables in Redis~\cite{redis} and balanced trees~\cite{DBLP:conf/sigmod/BangOMPB20,DBLP:conf/adc/LuNT00} often encounter bottlenecks due to lock contention and rebalancing overhead when handling numerous simultaneous time-series insertions and complex feature computations. Redis's hash tables also perform periodic rehashing to expand or shrink the table size, which can introduce significant latency spikes and impede real-time performance. To address these challenges, we adopt the refined skiplist data structure~\cite{DBLP:journals/cacm/Pugh90}, where nodes in the first layer are sorted by key values (e.g., item ID), and each key node in the second layer points to a linked list (or a secondary skiplist) that contains all data tuples associated with that key ordered by timestamps. 




\hi{Lock-Free Concurrent Access.} We support lock-free reads and writes over the refined data skiplist (with multiple levels of linked lists that can perform insertions and deletions independently) by utilizing atomic operations for pointer updates. Specifically, when a thread inserts or deletes a node, it uses atomic compare-and-swap (CAS) instructions to update pointers, i.e., a memory location is updated only if it contains an expected value. For instance, during insertion, the thread identifies the position where the new node should be placed and attempts to update the next pointer of the preceding node using CAS. If another thread has modified the pointer in the meantime, the CAS operation fails, and the thread retries the operation. 



\hi{Out-of-Date Data Removal.} Managing outdated data efficiently is essential for maintaining optimal performance and ensuring that \oursys does not process stale information that has surpassed its time-to-live (TTL). Here, we provide an efficient outdated data removal strategy: \bfit{(1) Timestamp Ordering:} By ordering nodes based on their timestamps, the skiplist allows quick identification of outdated tuples. Since the nodes are sorted, all outdated nodes are located contiguously at the beginning or end of the list, depending on the sorting order. \bfit{(2) Batch Deletion:} To remove all data before a specific timestamp \( T \), we traverse the skiplist from the head node and delete nodes until it reaches a node with a timestamp \( \geq T \). 

\subsection{On-Disk Data Structure}

For online data requiring persistence, we adopt RocksDB as the on-disk storage engine, used in databases like Redis~\cite{redis} and TiDB~\cite{tidb}. First, we assign multiple indexes when creating tables for columns frequently used in operations like order-by and partition-by. Each index corresponds to a Column Family in RocksDB. Different column families have separate Sorted String Table (SST) files and individual data eviction policies, but they share a single Memory Table (memtable). This means that while each Column Family manages its on-disk storage and data eviction independently, they all write to the same in-memory data structure before flushing data to disk.  In our implementation, we utilize the refined data skiplist as the Memory Table, where a key and a timestamp (ts) are concatenated to form a composite key. The RocksDB keys are pre-sorted, so data with the same key is grouped together, facilitating data queries over a specified time range. We support data eviction in RocksDB by parsing the keys and timestamps in the skiplist and removing data whose timestamp is out-of-date.

\vspace{-.5em}
\section{Memory Management Mechanisms}
\label{sec:resource}

\subsection{Table-Level Memory Configuration}

\oursys workloads often require substantial memory resources, particularly when dealing with large-scale feature computations. To guide users in choosing appropriately sized memory configurations, we provide an empirical memory estimation model.

\hi{Memory Usage Estimation.} Inspired by established practices in feature computation~\cite{openmldbblog}, we use the following model to estimate total memory usage:
\begin{align*}
\text{mem\_total} 
&= \sum_{i=1}^{n_{\text{table}}} n_{\text{replica}_{i}} \biggl[\, \biggl(\sum_{j=1}^{n_{\text{index}}} n_{\text{pk}_{ij}} \times (\lvert pk_{ij}\rvert + 156)\biggr) \\
&\quad +\, n_{\text{index}_{i}} \times n_{\text{row}_{i}} \times C \;+\; K \times n_{\text{row}_{i}} \times \lvert row_{i}\rvert \biggr].
\end{align*}
where $n_{\text{table}}$ is the total number of tables, and $n_{\text{index}}$ is the number of indexes per table. For each table $i$, $n_{\text{replica}_{i}}$ is the number of replicas (full table data copies), $n_{\text{pk}_{ij}}$ is the number of unique primary keys on the $j$th index column, and $|pk{ij}|$ is the average key column length. The parameters $n_{\text{index}_{i}}$ and $n_{\text{row}_{i}}$ denote the number of indexes and tuples in table $i$, respectively, while $|row_{i}|$ is the average tuple length. The factor $C$ depends on the table type: set $C=70$ for ``latest'' (storing recent data for a given key) or ``absorlat'' (with additional logic like combining recent entries) tables, and $C=74$ for ``absolute'' (data keyed by a timestamp or an absolute reference) or ``absandlat'' (e.g., data accessible by absolute timestamps) tables. The parameter $K$, representing the number of data copies stored, varies between 1 and $n_{\text{index}}$.  For instance, suppose we have a ``latest'' table with 1 million rows, average row length 300 bytes, two indexes, and replicas equal 2. Each primary key is 16 bytes. Using $C=70$ and $K=1$, the memory usage is about 1.568 GB.  Note memory consumption by pre-aggregation are typically negligible compared to the main memory consumption and so is not reflected here.



With the estimated memory usage, users can flexibly assign storage engines to tables in \oursys. For instance, if the estimated memory usage for a table fits within available memory resources and ultra-low latency (around 10 milliseconds) is required, the in-memory storage engine is recommended. If memory resources are limited or the estimated memory usage exceeds available memory, and a latency of 20-30 milliseconds is acceptable, the disk-based storage engine may be considered, leading to approximately 80\% savings in hardware costs (see Section~\ref{subsec:performance}).

\vspace{-.5em}
\subsection{Runtime Memory Management}

Moreover, to ensure high stability and prevent memory exhaustion, which can lead to tablets being terminated by the operating system and result in service downtime, \oursys introduces two key features: \bfit{(1) Memory Resource Isolation} provides a tablet-level configuration parameter (\texttt{max\_memory\_mb}) that limits the maximum memory usage of a tablet. When memory usage exceeds this limit, write operations fail but read operations continue, keeping the service online. Users can address memory shortages through scaling, shard migration, or other methods with minimal impact on production systems. \bfit{(2) Memory Alerting Mechanism:} An alerting mechanism notifies relevant personnel or systems when memory usage surpasses a preset threshold. 

\section{Evaluation}
\label{sec:experiments}

\subsection{Experimental Setup}

\begin{sloppypar}
\hi{{Workloads}.} We evaluate \oursys using a micro-benchmark (\microbench) and various real-world production workloads. The micro-benchmark measures feature computation performance prior to deploying~\cite{openmldbbench}. For real workloads, we consider the publicly available TalkingData dataset from Kaggle~\cite{talkdata}, as well as two production use cases at Akulaku~\cite{akulaku}: an Item Ranking service (\realbenchone) and a Geographical Location Querying service (\realbenchtwo). 
\end{sloppypar}

\noindent$\bullet$ \bfit{\microbench.} We implemented a Java-based testing tool employing the \oursys Java SDK and Jedis to compare performance and memory consumption. Due to differences in supported data types and storage layouts, the data insertion procedure for each system was slightly adjusted. To reflect realistic usage, we designed time-series-based tests (with three stream tables) with multiple adjustable parameters (e.g., number of windows, number of LAST JOIN operations).

\noindent$\bullet$ \bfit{\membench.}  The open TalkingData dataset (covering around 200 million clicks over four days~\cite{talkdata}) contains common column types (e.g., strings, number values, timestamps). Since the dataset is time-series-based and many data tuples share the same \texttt{ip} key, we created a table indexed by the \texttt{ip} column, allowing us to calculate the memory usage savings by \oursys and the baselines.

\noindent$\bullet$ \bfit{\realbenchone.} The RTP service at Akulaku is an item-ranking framework with a data warehouse exceeding 1,000 TB. It supports various online machine learning models, such as Logistic Regression (LR) and Gradient Boosted Decision Trees (GBDT), as well as diverse feature formats. This framework powers multiple online ML tasks, including user risk ranking, device fingerprinting, and facial recognition.

\noindent$\bullet$ \bfit{\realbenchtwo.} The \realbenchtwo workload consists of 10 billion geospatial tuples linked to GPS coordinates. It involves complex, full-scale geospatial queries (e.g., proximity searches, route optimization, clustering), which are extremely resource-intensive and can cause out-of-memory (OOM) errors in traditional systems.

\hi{Test Settings.} For \microbench, we use four servers of identical configurations: each with 512GB of memory, dual Intel Xeon E5-2630 v4 CPUs (2.20GHz), a 1Gbps network card, and running CentOS-7. Three servers run the \oursys service, and the fourth hosts the client application using the \oursys Java SDK for testing. For other workloads, experiments are conducted on 16 servers of similar configurations.

\subsection{Overall Performance}
\label{subsec:performance}

\subsection*{9.2.1 Online Feature Computation}
\label{subsec:online}

\begin{figure}[!t]
  \centering
  \includegraphics[width=\linewidth]{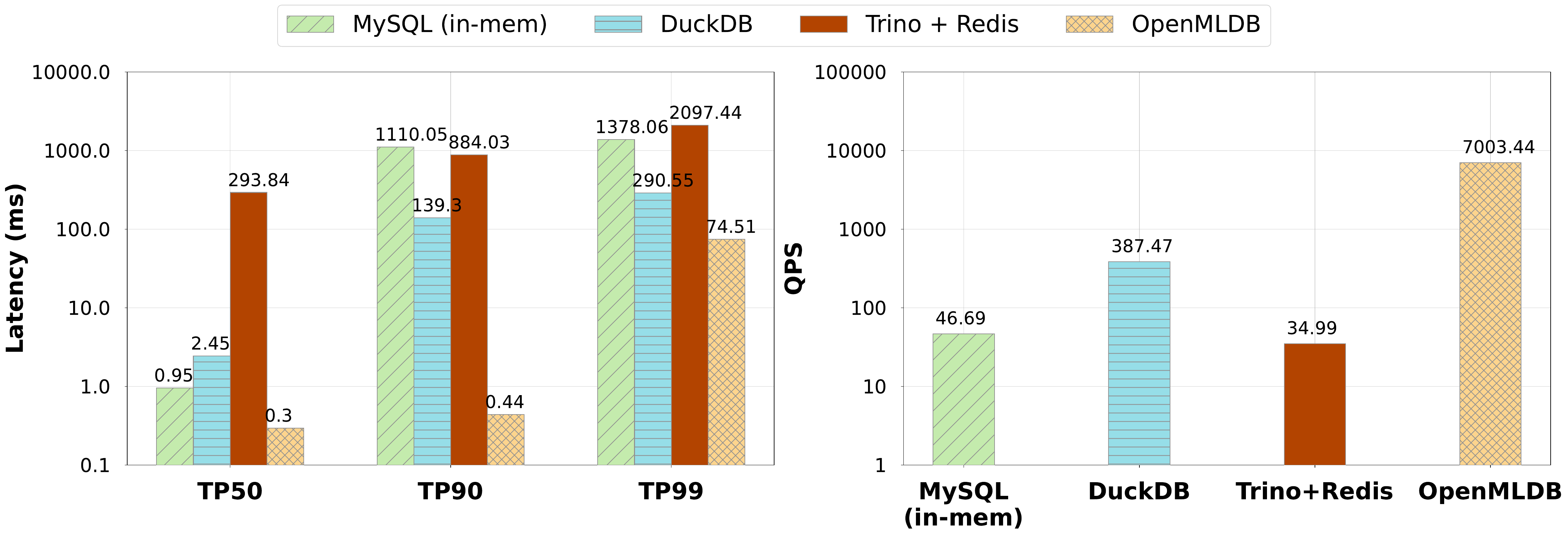}
  \vspace{-1.25em}
  \caption{Online \microbench Performance Comparison.}
  \label{fig:online-latency-res}
  \vspace{-.75em}
\end{figure}

\begin{figure}[!t]
  \centering
  \includegraphics[width=.75\linewidth]{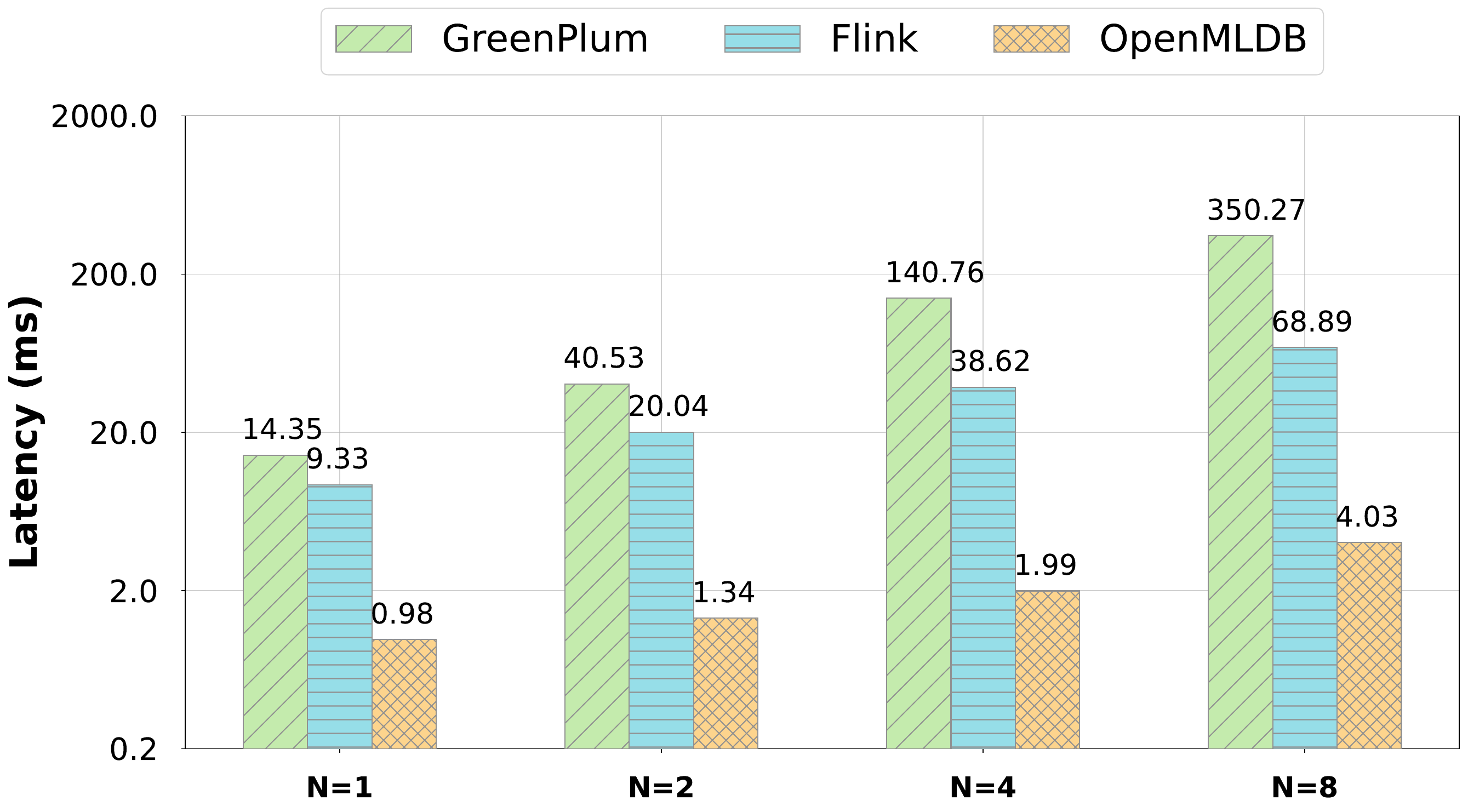}
  \vspace{-.5em}
  \caption{\realbenchone Performance Comparison.}
  \label{fig:online-real-exp}
  \vspace{-1.5em}
\end{figure}

\begin{sloppypar}
\hi{\microbench Performance.} We first compare \oursys with three commonly used baselines for in-memory online analytics: (1) \trino: We pair Redis (a popular in-memory data store~\cite{redis}) with Trino (an ANSI SQL-compliant engine~\cite{trino}) to handle SQL queries over Redis data. (2) \mysql: We test MySQL configured with the MEMORY storage engine to compute online features over the stored relational data~\cite{mysql}. (3) \duckdb: We also compare \duckdb, an embedded analytical database~\cite{duckdb} designed for online analytical processing. As shown in Figure~\ref{fig:online-latency-res}, \oursys achieves significant performance improvement in both latency and throughput compared to all the baselines on \microbench. In terms of latency, \oursys outperforms \mysql by over 68.4\%, \duckdb by 87.7\%, and \trino by over 96\%, respectively. \oursys also achieves throughput gains by over \emph{17 times} higher than the baselines. The reasons are three-fold. 
\end{sloppypar}

\noindent$\bullet$ \oursys employs a C++-based compilation framework that specializes window operations across multiple granularities (e.g., daily, weekly, and monthly) in a single integrated pass. \oursys also uses LLVM-based Just-In-Time (JIT) compilation to fuse and streamline aggregate functions (e.g., sum and avg). Instead, \mysql relies heavily on interpreted SQL execution, and \duckdb does not incorporate aggressive cross-operation optimizations. Moreover, compared to \trino’s Java framework with frequent RPC calls, \oursys benefits from more direct control over low-level operation optimizations and can apply advanced caching strategies that further reduce overheads for reusable feature scripts.

\noindent$\bullet$ \oursys leverages the in-memory data structure to pre-rank tuples by timestamps. As a result, data retrieval and scanning occur in a nearly linear time complexity with respect to input sizes. In contrast, \mysql often incurs overhead from disk-based indexing, while \duckdb, though in-memory and columnar, may still require additional passes for complex temporal queries. \trino, on the other hand, spreads window-state management over multiple operators, introducing extra latency for state lookups and aggregation steps.

\noindent$\bullet$ \oursys dynamically redistributes data by keys and employs incremental computation strategies for windowed operations. Instead of reprocessing entire datasets for each new computation like  \mysql and \duckdb, \oursys updates only the necessary segments of the data stream. Likewise, compared to \trino’s distributed streaming model, which often involves shuffling and checkpointing overhead, \oursys’s key-based redistribution and incremental approaches minimize redundant computations, thereby significantly reducing overall processing time and improving real-time responsiveness.

\hi{\realbenchone Performance.} The Akulaku company has integrated \oursys into its ML pipeline to streamline real-time feature generation and ranking~\cite{akulaku}. Figure~\ref{fig:online-real-exp} shows the real performance tested in Akulaku's RTP business, which compares \oursys with \greenplum (a typical MPP database~\cite{greenplum}) and the streaming system \flink~\cite{apacheflink}. We can see \oursys significantly outperforms the baselines for real-time TopN queries. \greenplum incurs prohibitive recomputations for new data tuples, causing substantial delays. \flink, although good at processing basic time windows, is also not well optimized for TopN ranking. In contrast, {\it \oursys scales nearly linearly}, achieving Top1 queries in about 0.98 ms and Top8 queries around 5 ms, significantly better than \flink’s sub-100 ms range. The improvement is due to (1) \oursys’s runtime dynamic compilation and LLVM IR code generation, which embed query parameters directly into machine code to improve execution efficiency without sacrificing generality, and (2) \oursys's in-memory data structure pre-ranks stream data by keys and timestamps, thereby minimizing runtime sorting overhead.

\begin{table}[!t]
\centering
\caption{Memory resource saved by \oursys (bytes).}
\vspace{-1em}
\begin{tabular}{|r|r|r|r|}
\hline
\textbf{\#-Tuples} & \textbf{RedisMem} & \textbf{\oursys Mem} & \textbf{{Reduction}} \\
\hline
10,000 & 9,272,328 & 2,339,699 & $\downarrow$ 74.77\% \\
100,000 & 48,501,288 & 15,624,290 & $\downarrow$ 67.79\% \\
1,000,000 & 215,323,024 & 105,722,441 & $\downarrow$ 50.90\% \\
10,000,000 & 1,897,343,984 & 1,008,276,458 & $\downarrow$ 46.86\% \\
184,903,890 & 34,071,049,864 & 18,513,271,540 & $\downarrow$ 45.66\% \\
\hline
\end{tabular}
\label{table:memory_comparison}
\vspace{-1em}
\end{table}

\begin{sloppypar}
\hi{Memory Consumption.} We compare memory usage between \oursys and \trino. As shown in Table~\ref{table:memory_comparison}, we can see \oursys achieves over 50\% memory saving than \trino, which can be extremely vital for practical usage and saving machine costs. Specifically, for smaller datasets (10K tuples), \oursys uses 74.77\% less memory than \trino. As dataset size grows to hundreds of millions of tuples, the memory reduction stabilizes around 45.66\%. This memory efficiency stems from \oursys’s compact time-series encoding and specialized indexing, reducing overhead from repeated keys and non-compact data layouts typical in \trino or other general-purpose stores. By minimizing memory usage, \oursys not only lowers hardware costs but also processes more data in-memory, enabling more scalable and lower-latency feature computations than baseline systems that must either add more hardware or spill to disk.
\end{sloppypar}

\subsection*{9.2.2 Offline Feature Computation}
\label{subsec:offline}

\begin{figure}[!t]
  \centering
  \includegraphics[width=\linewidth]{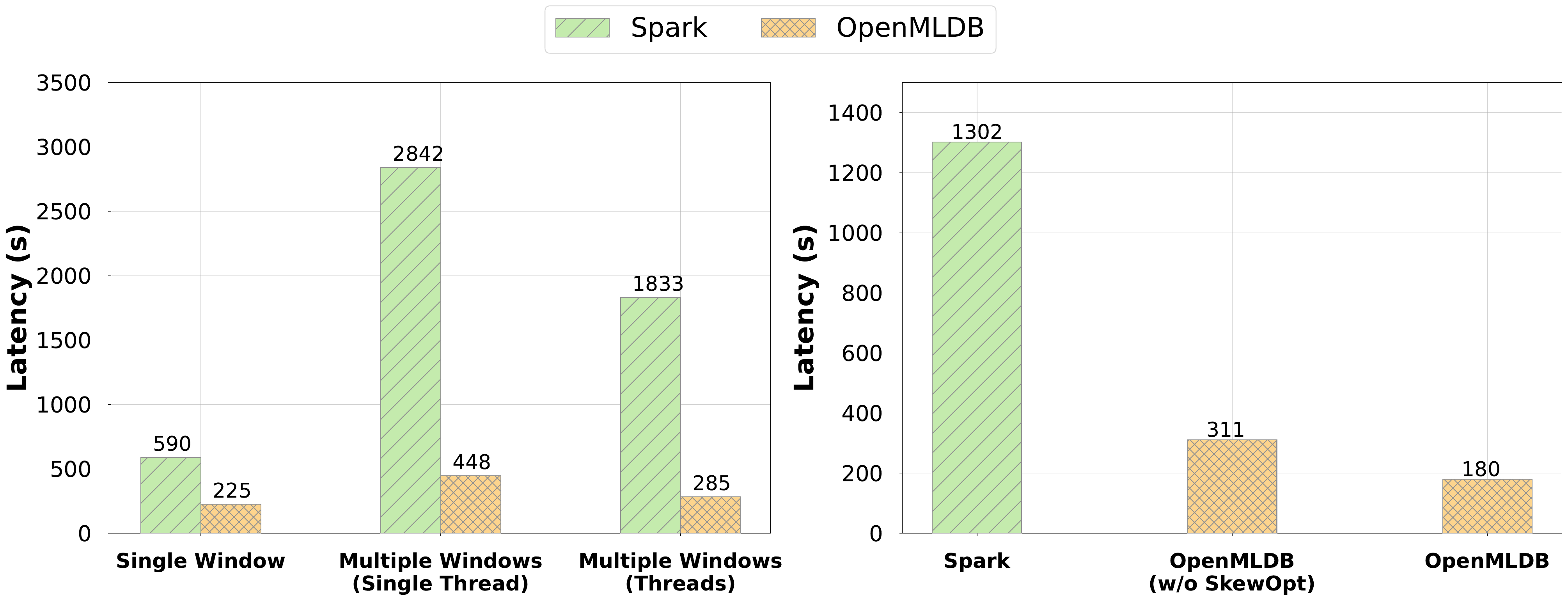}
  \vspace{-1.25em}
  \caption{Offline \microbench Performance Comparison.}
  \label{fig:performance-exp}
  \vspace{-.75em}
\end{figure}

\hi{\microbench Performance.} As shown in Figure~\ref{fig:performance-exp}, on the offline \microbench, \oursys outperforms Spark significantly, delivering a 2.6x speedup on single-window queries, a 6.3x speedup on multiple-window workloads. Against skewed data (imbalances across the distributed data partitions), \oursys reduces latency by 7.2x with skew optimization (180s vs. 1302s). The reasons are three-fold. First, \oursys employs C++ execution libraries and leverages just-in-time (JIT) compilation to generate efficient native machine code. In contrast, Spark’s query execution relies heavily on the JVM runtime, resulting in additional overhead from garbage collection and interpretation. Second, \oursys prioritizes direct in-memory window operations to reduce CPU cache misses. 
It efficiently parallelizes time-window computations and streamlines memory usage.
While Spark must orchestrate multiple stages and shuffles, incurring expensive serialization, deserialization, and data movement.  
Third, \oursys removes unnecessary initialization routines and repetitive object constructions. By trimming down code paths and utilizing a more straightforward execution model, \oursys enhances both performance and maintainability. In contrast, Spark’s more complex execution pipelines and object-heavy data structures introduce additional execution overhead.

\begin{figure}[!t]
  \centering
  \includegraphics[width=.6\linewidth]{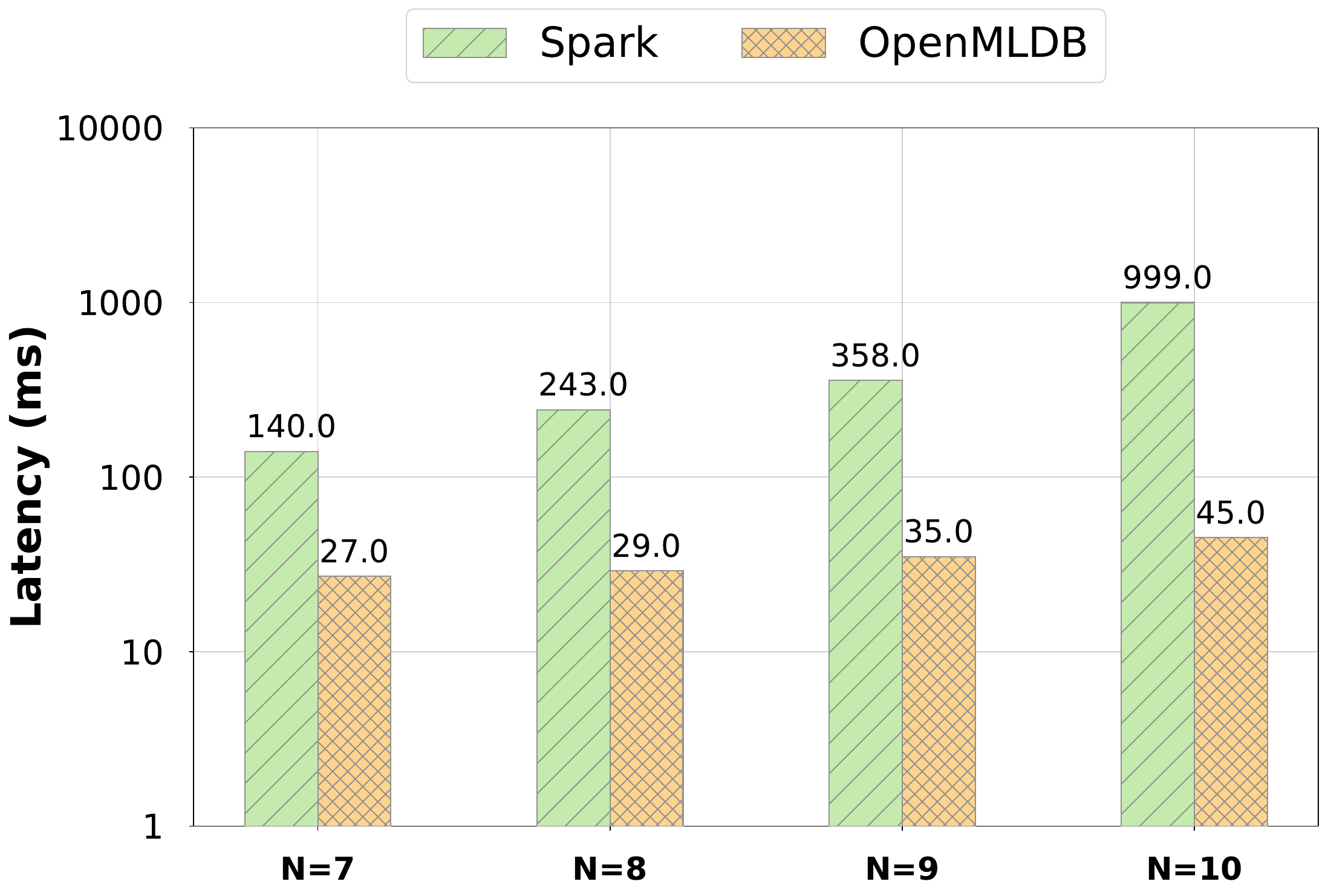}
  \vspace{-1.25em}
  \caption{Offline \realbenchtwo Performance Comparison.}
  \label{fig:offline-exp-real}
  \vspace{-.75em}
\end{figure}

\begin{sloppypar}
\hi{\realbenchtwo Performance.} Moreover, we compare the performance of \oursys and Spark in handling real geographic location scenarios involving complex queries. These queries necessitate evaluating the relative relationships among all GPS coordinates, thus requiring computation over the entire dataset for each query. This makes traditional SQL-based expressions inadequate and demands \oursys SQL. As shown in Figure~\ref{fig:offline-exp-real}, \oursys achieves near 30-millisecond response time, with execution speedups ranging from approximately 5x to over 22x compared to Spark as the hyper-parameter \emph{N} increases from 7 to 10. By leveraging LLVM-based compilation optimizations, \oursys generates more efficient execution code than Spark. Additionally, Spark’s memory structure exhibits inefficiencies in handling homogeneous, single-purpose tasks and encounters out-of-memory (OOM) errors for full-table queries. These observations underscore \oursys’s superiority in optimizing data formats and memory management.
\end{sloppypar}

\begin{figure}[!t]
    \centering
    \begin{subfigure}{0.24\textwidth}
        \includegraphics*[width=\textwidth, trim=10 10 10 50, clip]{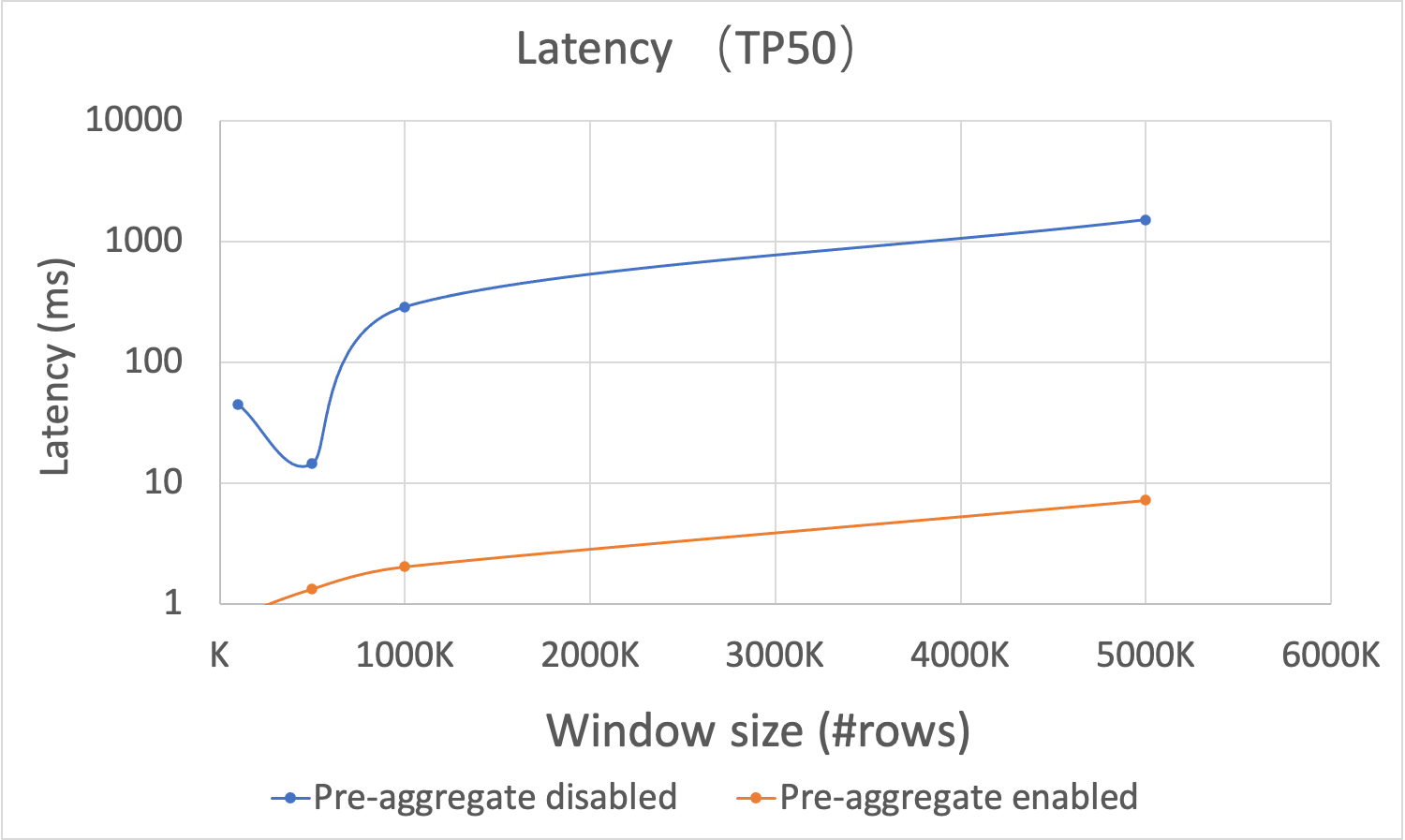}
        \caption{Latency}
        \vspace{6pt}
    \end{subfigure}
    \begin{subfigure}{0.22\textwidth}
        \includegraphics*[width=1.05\textwidth, trim=10 10 10 50, clip]{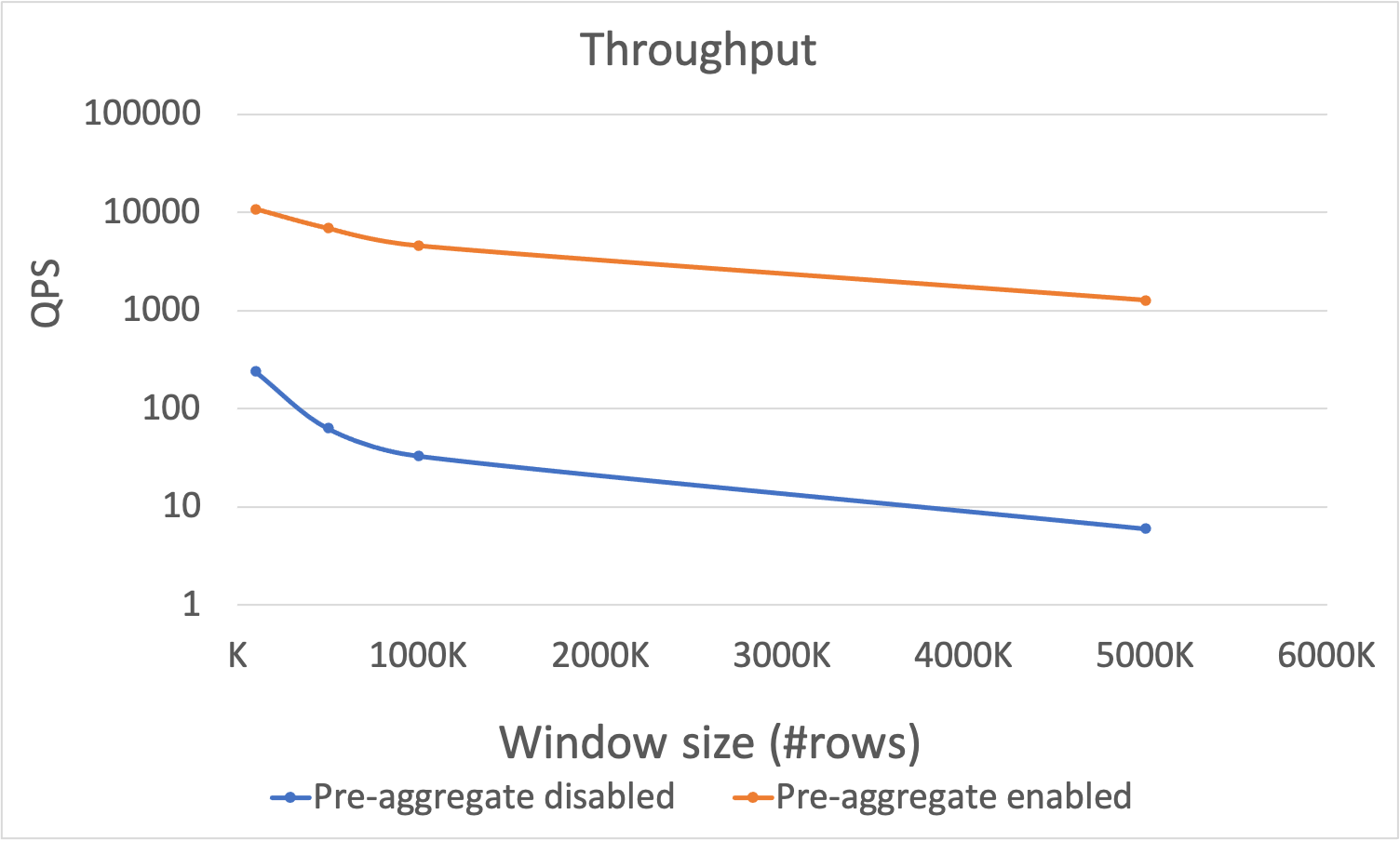}
        \caption{Throughput}
        \vspace{6pt}
    \end{subfigure}
    \vspace{-1.5em}
    \caption{Pre-Aggregation Performance.}
    \vspace{-1em}
    \label{fig:preaggregate-exp}
\end{figure}

\subsection{Ablation Study}
\label{sec:ablation}

\subsection*{9.3.1 Long Window Pre-Aggregation}
\label{subsec:preagg-exp}

To evaluate the impact of pre-aggregation on long windows, we configure a series of SQL queries with varying window sizes, ranging from 100K to 5000K tuples. As shown in Figure~\ref{fig:preaggregate-exp}, the performance improvement from enabling pre-aggregation is substantial. Without pre-aggregation, as the window size grows, \oursys must repeatedly scan and aggregate large volumes of raw data, leading to exponential growth in latency and a precipitous drop in throughput. In contrast, with pre-aggregation enabled, intermediate partial aggregates are precomputed and cached, drastically reducing the overhead for subsequent queries over the same window.

\begin{figure}[!t]
    \centering
    \begin{subfigure}{0.25\textwidth}
        \includegraphics*[width=\textwidth]{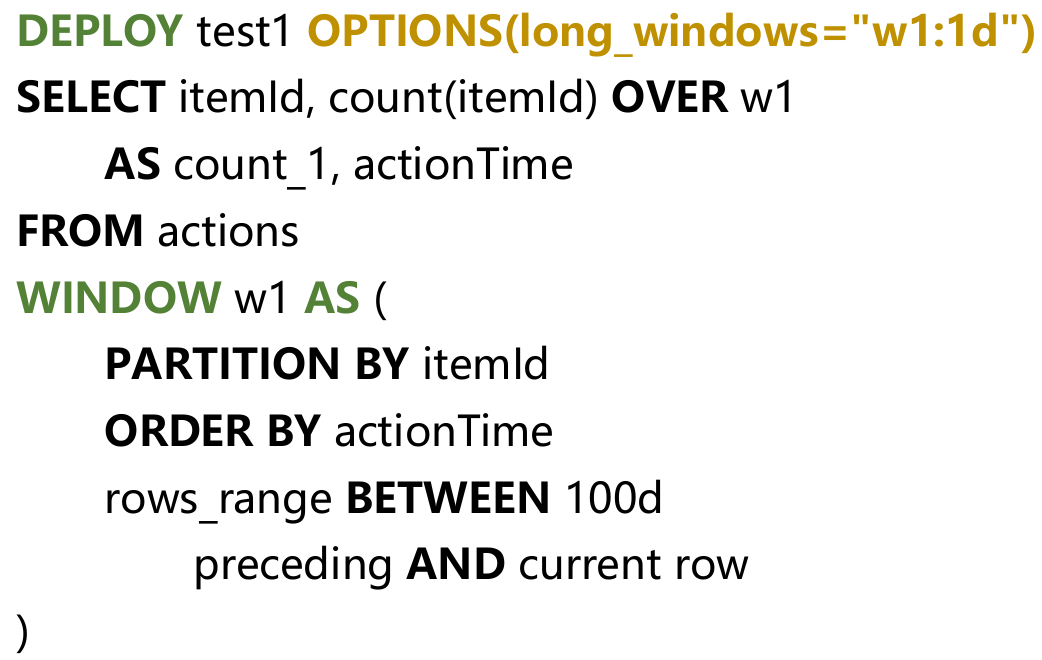}
        \caption{Long Window SQL}
        \vspace{3pt}
    \end{subfigure}
    \begin{subfigure}{0.2\textwidth}
        \includegraphics*[width=\textwidth]{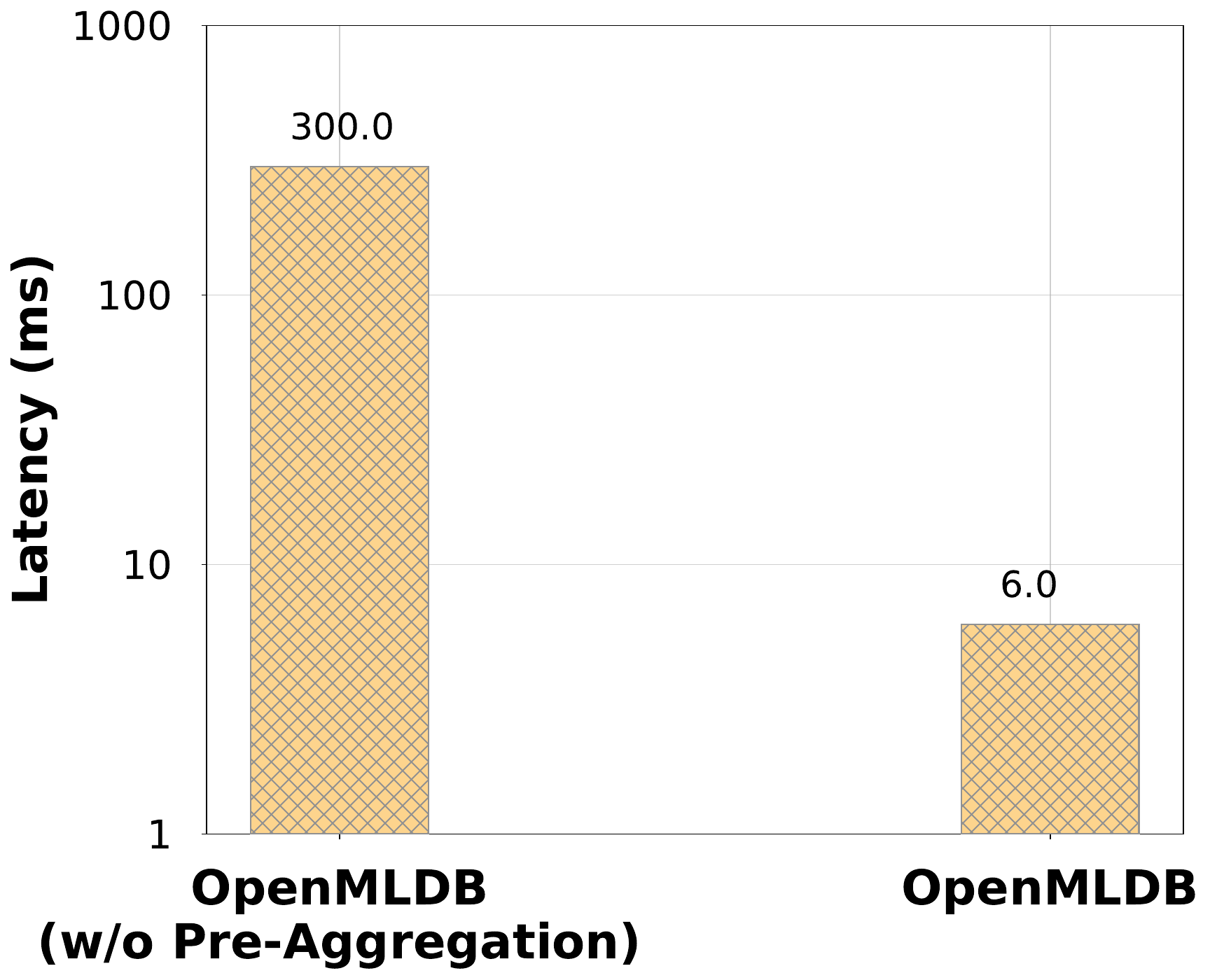}
        \caption{{Latency Comparison}}
        \vspace{3pt}
    \end{subfigure}
    \vspace{-1em}
    \caption{Performance of Long-Window Optimization: (a) Testing Feature Script and (b) Latency Comparison.}
    \vspace{-.75em}
    \label{fig:long_window}
\end{figure}

\begin{sloppypar}
Furthermore, we evaluate the effectiveness of long window pre-aggregation over a larger dataset of 860,000 tuples. As shown in Figure~\ref{fig:long_window}, we observe that the long window pre-aggregation in \oursys significantly improves the efficiency of a request with little SQL modification and a slightly higher data loading overhead. Specifically, when specifying the usage of the long window pre-aggregation, i.e., adding \textsf{\emph{deploy test1 OPTIONS(long\_windows="w1:1d")}} in the SQL, the response time for a request obtains a $45 \times$ reduction (i.e., from $300ms$ to $6ms$) with this optimization technique. Therefore, while pre-aggregation introduces a small overhead during data loading, \oursys's online computation efficiency can be greatly improved, especially for queries with repeated window aggregations. 
\end{sloppypar}

\begin{figure}[!t]
  \centering
\includegraphics[width=.75\linewidth]{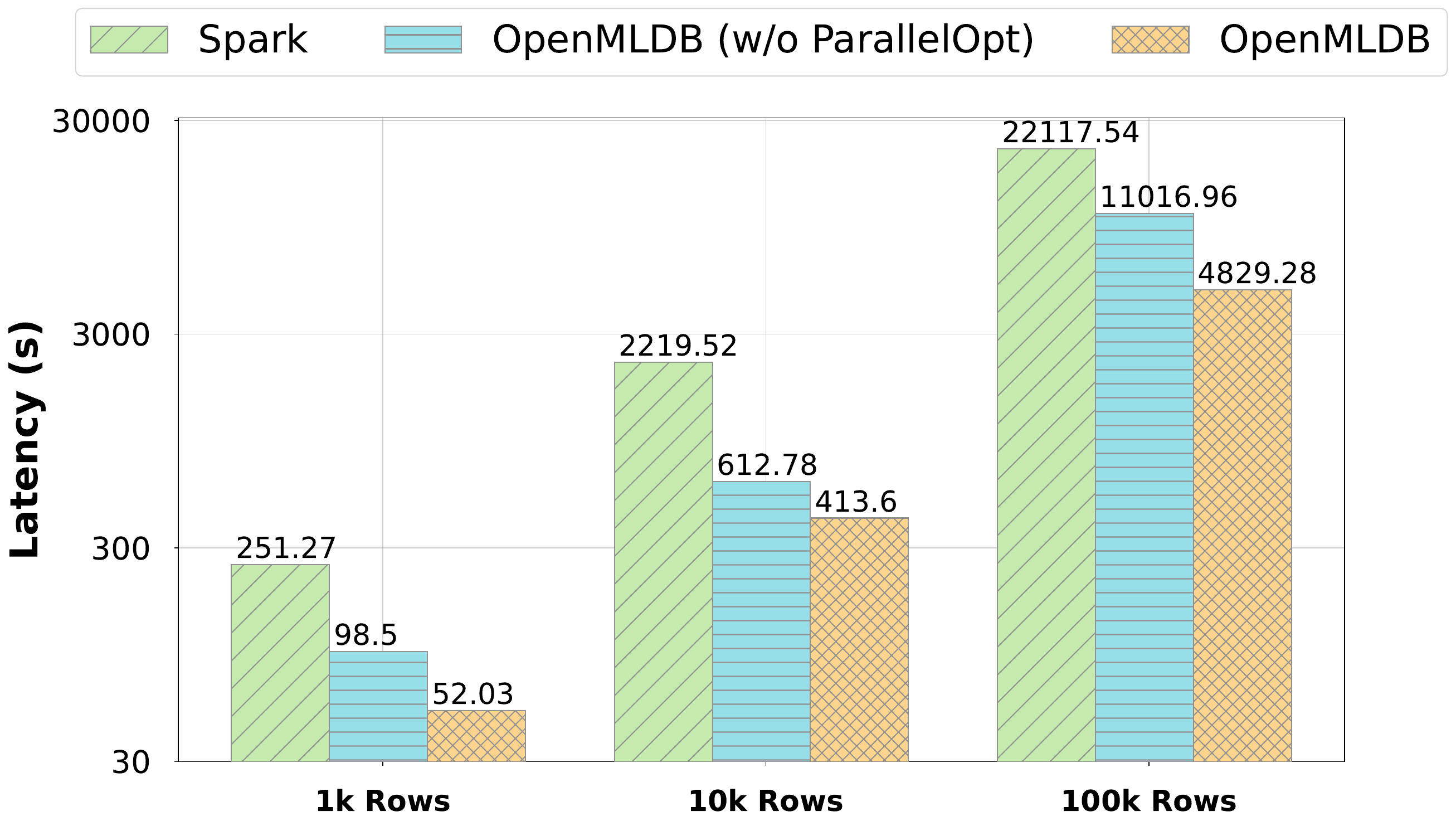}
  \vspace{-.75em}
  \caption{Multi-Window Parallel Optimization Performance.}
  \label{fig:parallel-exp}
  \vspace{-1.25em}
\end{figure}

\subsection*{9.3.2 Multi-Table Window Union Optimization}
\label{subsec:intervaljoin-exp}

Next, we evaluate the performance of multi-table window union operations. Without the self-adjusting techniques introduced in Section~\ref{subsec:interval-join}, throughput severely degrades as window sizes increase. For example, the static execution approach (e.g., the one in Apache Flink~\cite{apacheflink}) has a throughput of around 1000 tuples per second when the window size is 10,000 due to the need to reprocess extensive overlapping data intervals and handle skewed key distributions inefficiently. In contrast, \oursys maintains stable, high throughput (approximately 1,000,000 tuples/sec) across varying window sizes by leveraging its self-adjusting load balancing and incremental computation strategies. The dynamic key-to-thread mapping continuously redistributes workload to address shifting access patterns, mitigating hot spots that the static method struggles with.

\subsection*{9.3.3 Multi-Window Parallel Optimization}
\label{subsec:parallel-exp}

As shown in Figure~\ref{fig:parallel-exp}, we observe that \oursys with parallel optimization yields significant performance improvements. With small windows containing 1,000 rows, the performance of the smaller window becomes completely overshadowed in \oursys with parallel optimization. The overall task time perceived by the user is primarily the duration of the window with the longest execution time, resulting in a 4.8 times performance increase. In medium-sized windows, performance improves by 5.3 times, and parallel optimization on large windows yields a 4.6 times improvement compared to Spark. By concurrently processing multiple windows, \oursys spreads workloads across CPU cores more evenly and avoids serialization bottlenecks. Note while this parallel optimization in \oursys introduces overhead of some additional merging operations, we have minimized it through optimizations like the \textsf{Last Join} syntax for one-to-one correspondence with higher performance than traditional left joins (see Section~\ref{subsec:sql}).

\begin{figure}[!t]
  \centering
\includegraphics[width=.75\linewidth]{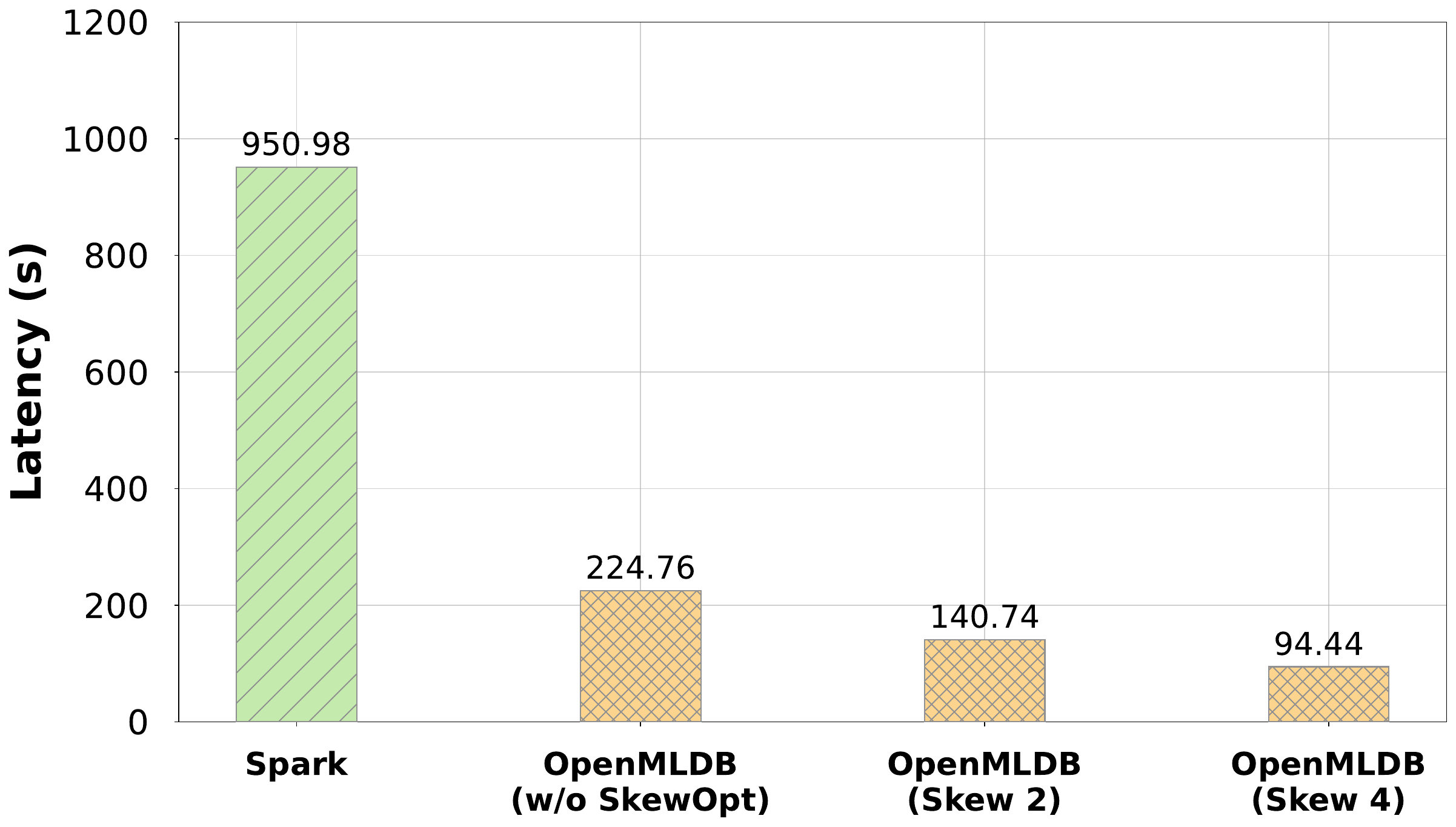}
  \vspace{-.5em}
  \caption{Data Skew Optimization Performance}
  \label{fig:data-skew-exp}
  \vspace{-1em}
\end{figure}

\subsection*{9.3.4 Data Skew Optimization}
\label{subsec:data-skew-exp}

Figure~\ref{fig:data-skew-exp} shows that, after applying data skew optimization, the performance has significantly improved compared to Spark and \oursys without skew optimization. Although \oursys without skew optimization already achieves 4 times performance improvement over Spark, the improvement becomes even greater with skew optimization, such as \textsf{skew 2} (the default value indicates the doubled data partition number). At \textsf{skew 4}, performance is 10.1 times higher over Spark and more than 2 times over without optimization. This improvement stems from the data-aware strategy that adjusts distributed partitions based on workload and data distributions. 

\subsection{Evaluation on Hyper-Parameters}
\label{subsec:hyper}

\begin{sloppypar}
Furthermore, we investigate how various parameters affect \oursys's performance, including \emph{(1) System parameters:} number of threads, whether to enable pre-aggregation optimization, \emph{(2) Query parameters:} number of windows, number of LAST JOINs, \emph{(3) Data parameters:} number of data tuples within a window, cardinality of the indexed column (i.e., the number of unique values after deduplication). Pre-aggregation optimization is usually only meaningful when the data volume within the window is huge (e.g., millions of data tuples), so it is turned off by default.
\end{sloppypar}

\begin{table}[h!]
\small
\centering
\caption{Performance for Different Feature Numbers.}
\vspace{-1.em}
\begin{tabular}{|c|c|c|c|c|c|c|}
    \hline
    \textbf{\#-Column} & \textbf{\#-Feature} & \textbf{TP50} & \textbf{TP90} & \textbf{TP95} & \textbf{TP99} & \textbf{TP999} \\
    \hline
    10 & 20 & 0.6 & 0.8 & 0.8 & 1.0 & 1.9 \\
    100 & 210 & 2.0 & 2.8 & 2.5 & 4.4 & 6.6 \\
    1000 & 2100 & 11.7 & 14.7 & 15.9 & 19.8 & 44.8 \\
    \hline
\end{tabular}
    \label{features-exp}
    \vspace{-.5em}
\end{table}

\hi{Number of Features.} To evaluate the scalability with respect to feature dimensions, we consider scenarios with varying numbers of columns and derived features. For instance, increasing from 10 columns (20 features) to 1,000 columns (2,100 features) leads to higher latency. However, the latency remains within tens of milliseconds, which is still acceptable for most online analytical workloads. As shown in Table~\ref{features-exp}, even at high feature counts, the observed latency is modest, confirming \oursys’s ability to handle complex feature pipelines without significant performance degradation.

\begin{figure}[!t]
    \centering
    \begin{subfigure}{0.24\textwidth}
        \includegraphics*[width=\textwidth, trim=10 10 10 50, clip]{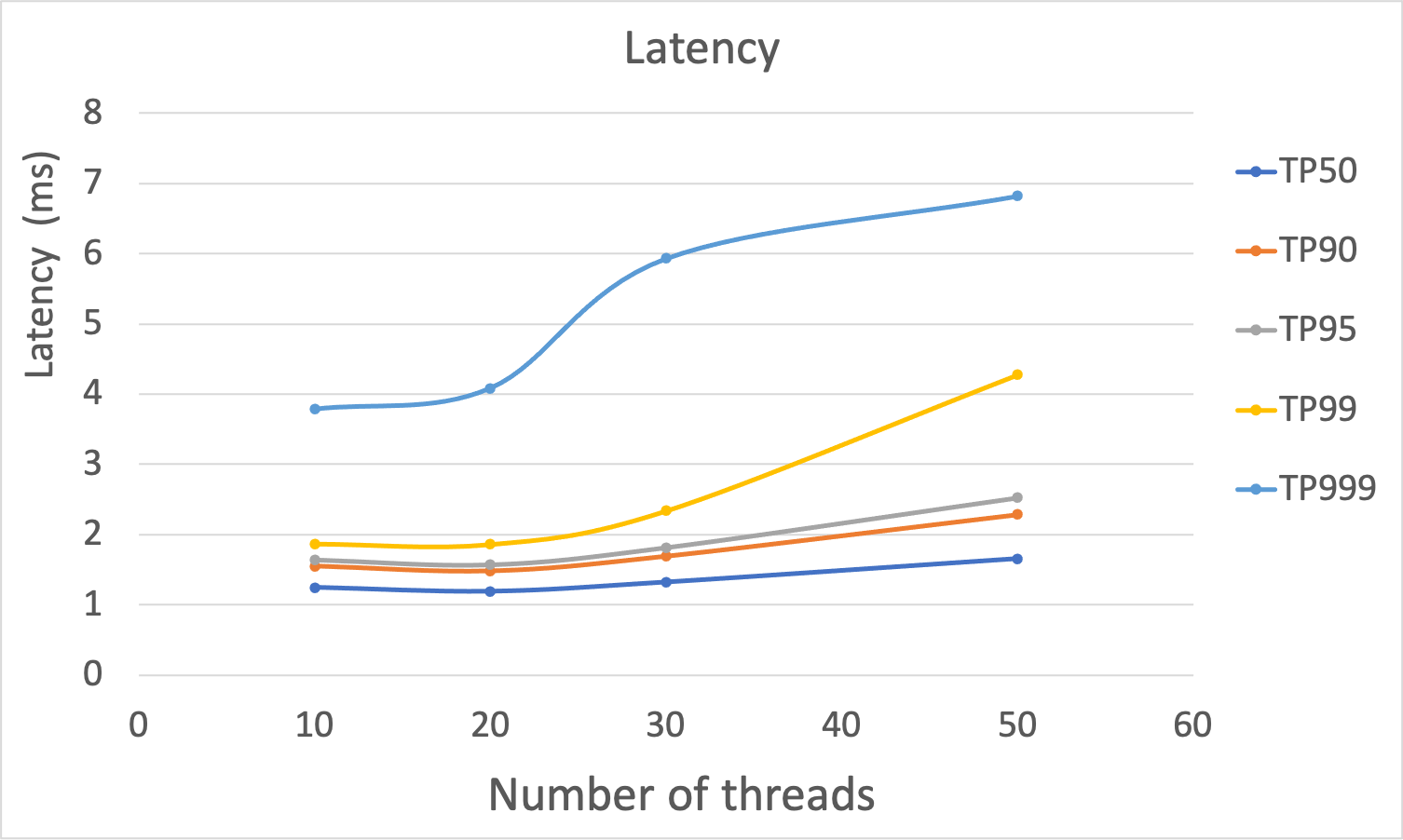}
        \caption{Latency}
        \vspace{3pt}
    \end{subfigure}
    \begin{subfigure}{0.22\textwidth}
        \includegraphics*[width=1.05\textwidth, trim=10 10 10 50, clip]{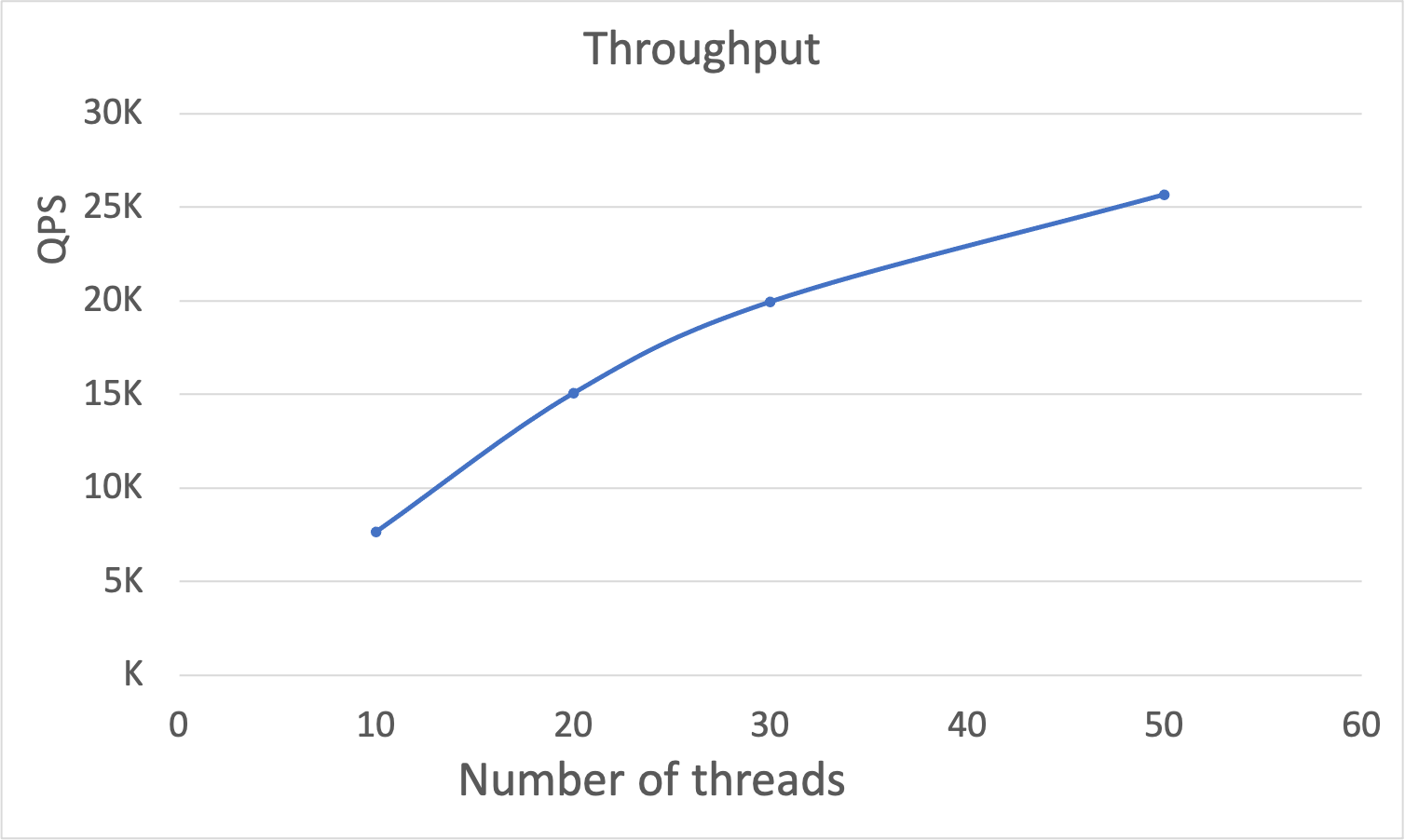}
        \caption{Throughput}
        \vspace{3pt}
    \end{subfigure}
    \vspace{-1.5em}
    \caption{Performance under Different Threads.}
    \vspace{-1em}
    \label{fig:threads-exp}
\end{figure}

\begin{figure}[!t]
    \centering
    \begin{subfigure}{0.24\textwidth}
        \includegraphics*[width=\textwidth, trim=10 10 10 40, clip]{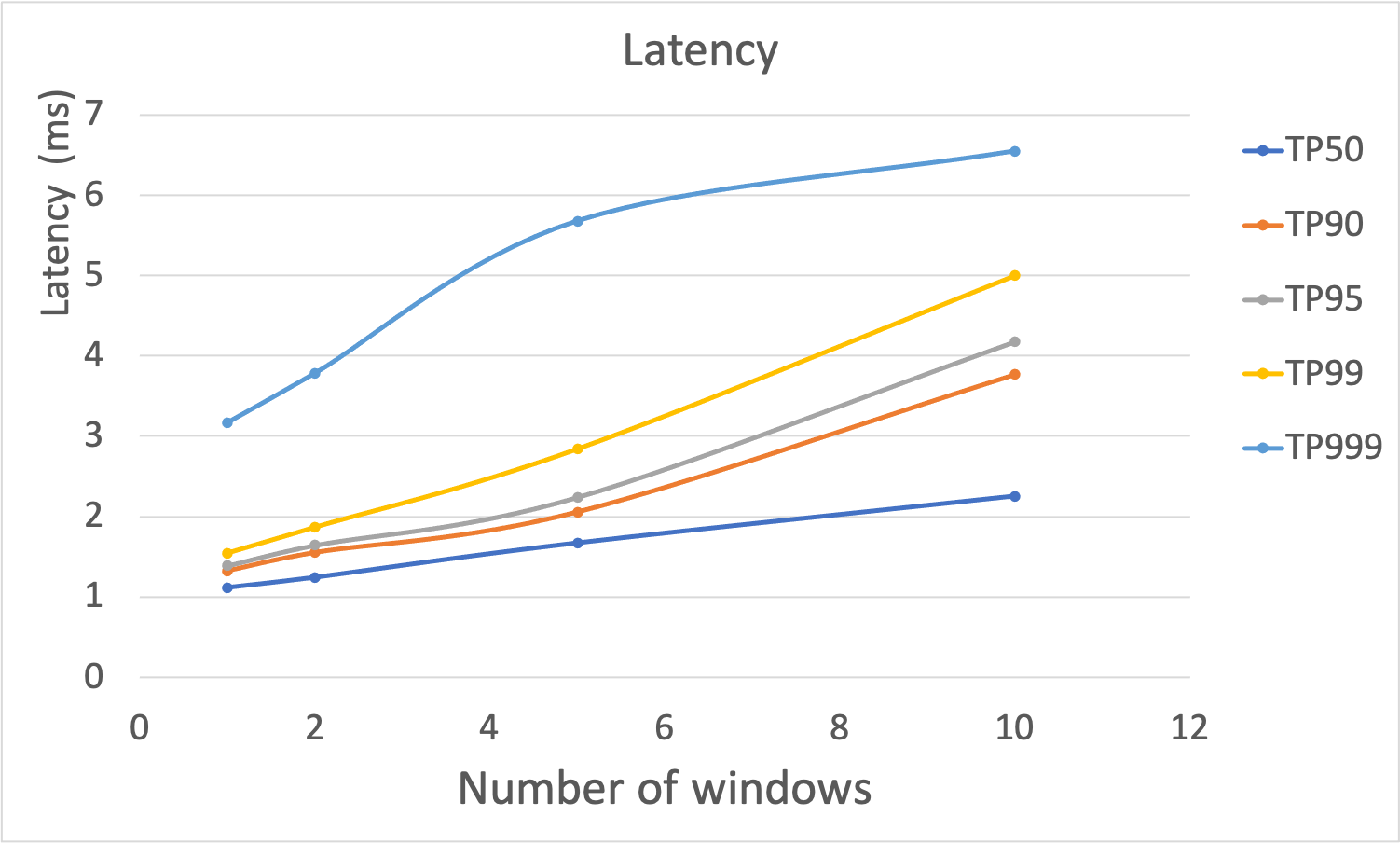}
        \caption{Latency}
        \vspace{3pt}
    \end{subfigure}
    \begin{subfigure}{0.22\textwidth}
        \includegraphics*[width=1.05\textwidth, trim=10 10 10 50, clip]{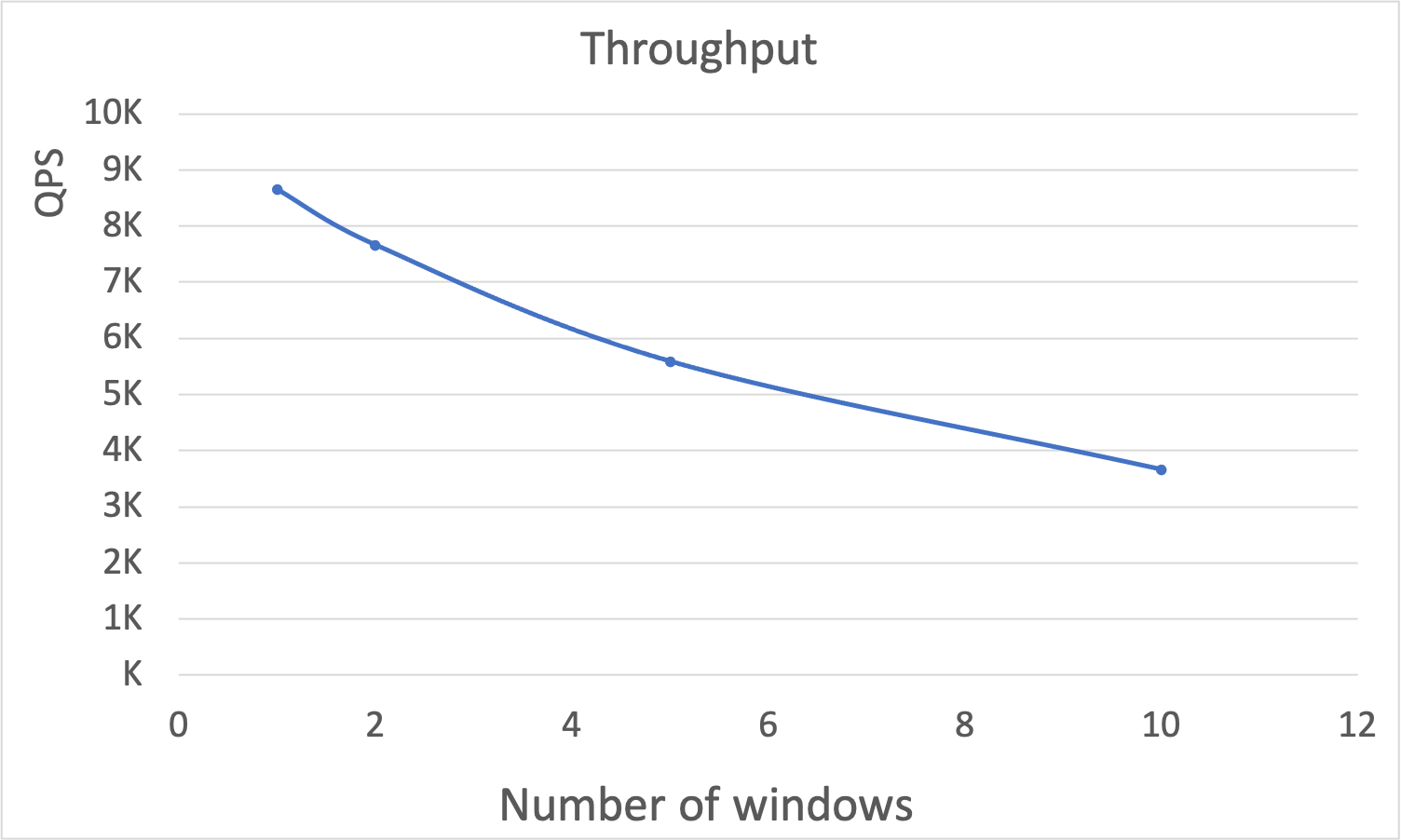}
        \caption{Throughput}
        \vspace{3pt}
    \end{subfigure}
    \vspace{-1.5em}
    \caption{Performance under Different Window Numbers.}
    \vspace{-1em}
    \label{fig:windows-exp}
\end{figure}

\begin{figure}[!t]
    \centering
    \begin{subfigure}{0.24\textwidth}
        \includegraphics*[width=\textwidth, trim=10 10 10 50, clip]{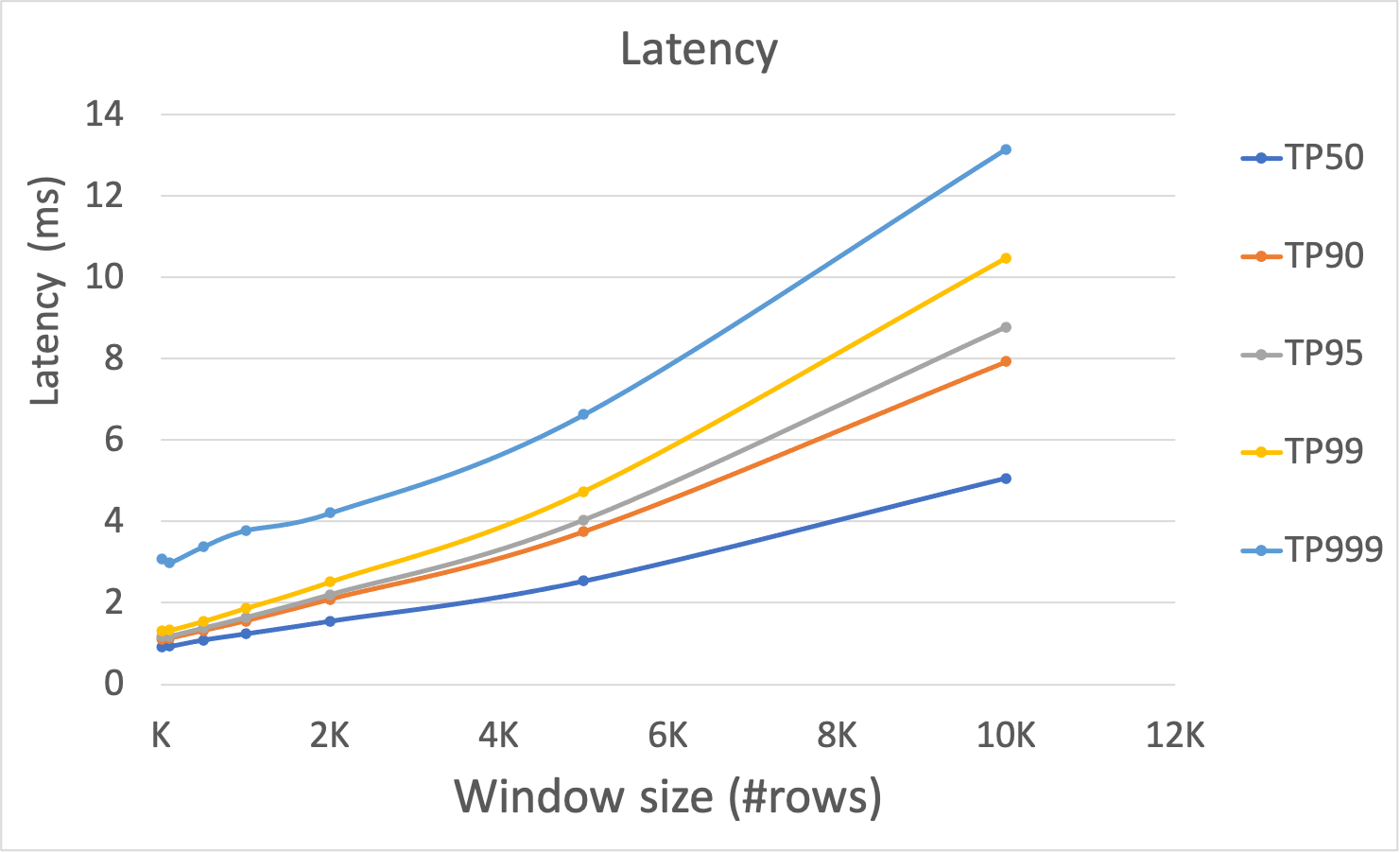}
        \caption{Latency}
        \vspace{3pt}
    \end{subfigure}
    \begin{subfigure}{0.22\textwidth}
        \includegraphics*[width=1.05\textwidth, trim=10 10 10 50, clip]{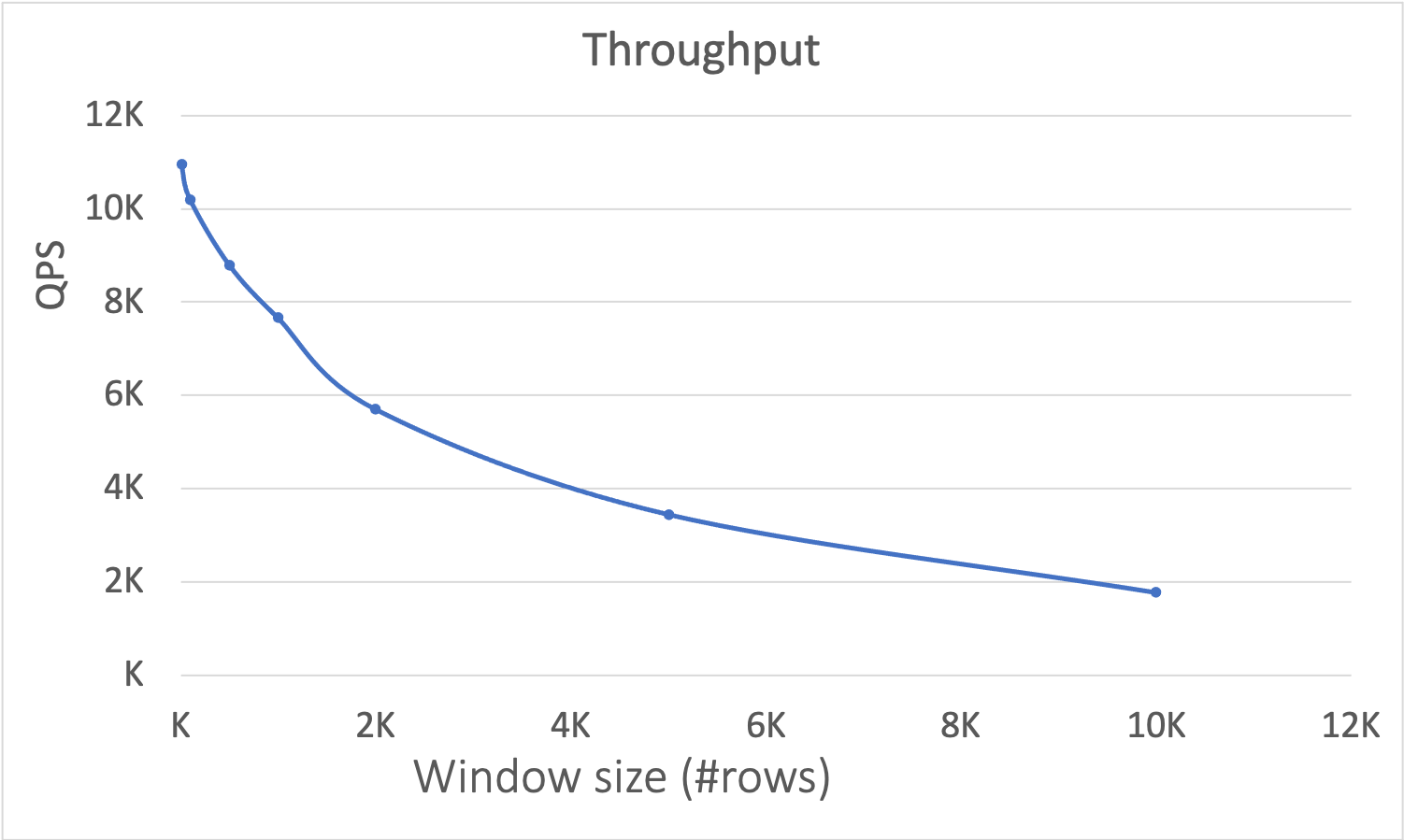}
        \caption{Throughput}
        \vspace{3pt}
    \end{subfigure}
    \vspace{-1.5em}
  \caption{Performance under Different Window Sizes.}
  \label{fig:tuplenumber-exp}
    \vspace{-1em}
\end{figure}

\begin{figure}[!t]
    \centering
    \begin{subfigure}{0.24\textwidth}
        \includegraphics*[width=\textwidth, trim=10 10 10 50, clip]{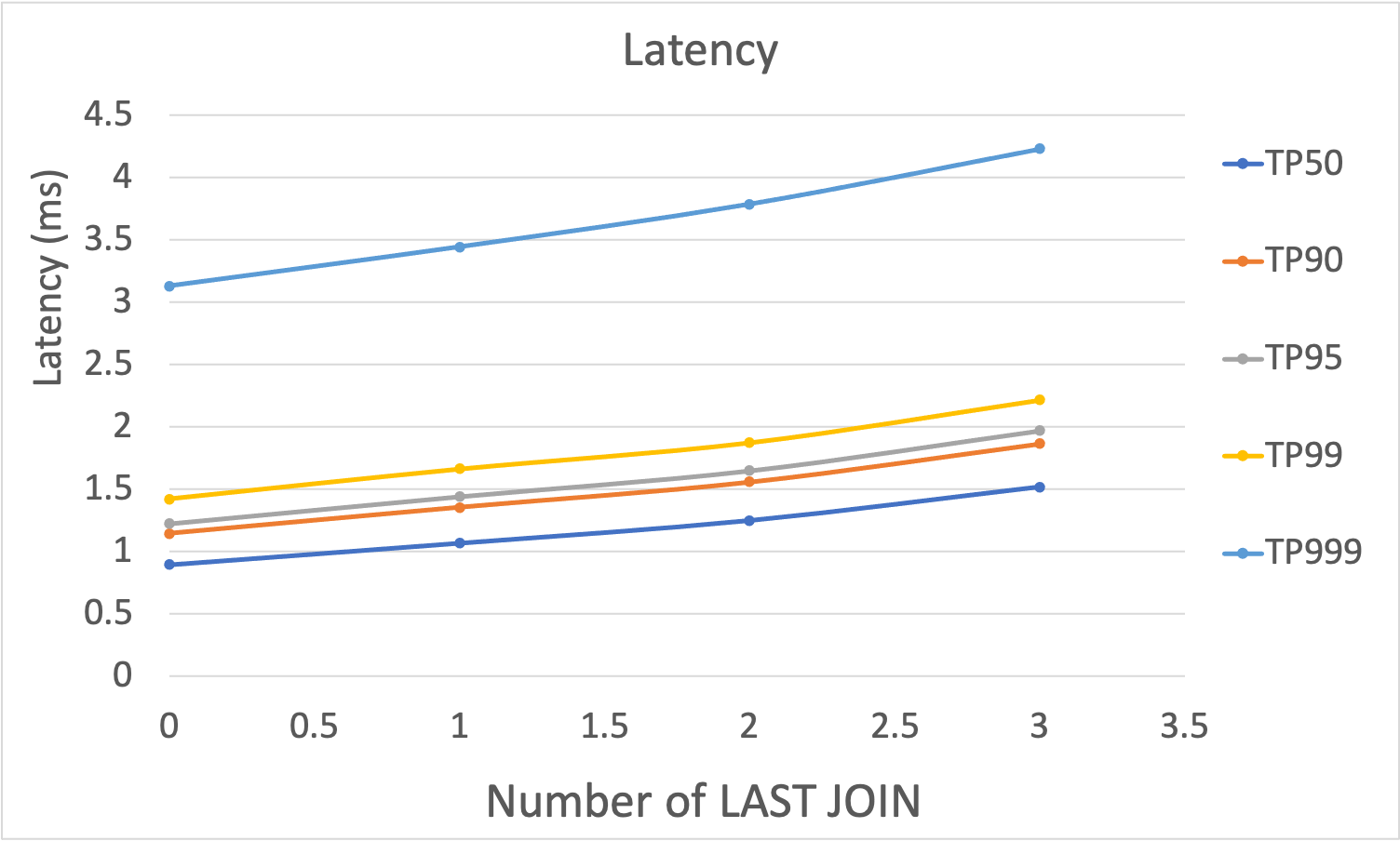}
        \caption{Latency}
        \vspace{4pt}
    \end{subfigure}
    \begin{subfigure}{0.22\textwidth}
        \includegraphics*[width=1.05\textwidth, trim=10 10 10 50, clip]{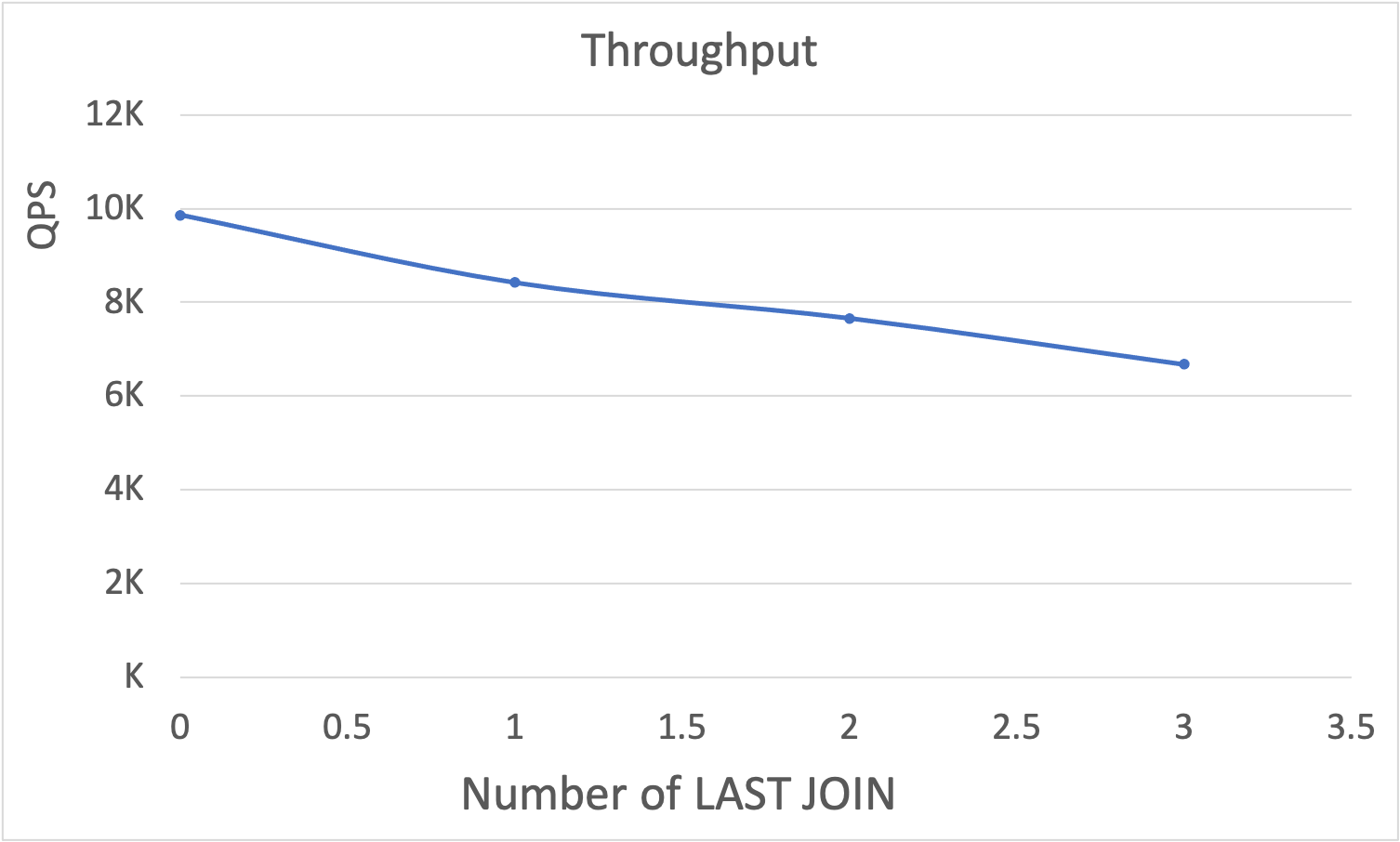}
        \caption{Throughput}
        \vspace{4pt}
    \end{subfigure}
    \vspace{-1.5em}
  \caption{Performance under Different Join Numbers.}
    \label{fig:joinnumber-exp}
    \vspace{-1.6em}
\end{figure}

\hi{Number of Threads.} Figure~\ref{fig:threads-exp} shows that increasing threads improves throughput but can slightly increase latency. Beyond 20 threads, latency grows but remains within single-digit milliseconds. This trade-off allows users to tune resource usage for their latency vs. throughput requirements. Such scaling behavior is more predictable and manageable than in Spark or MySQL, where concurrency often introduces non-linear slowdowns or GC overhead.

\hi{Number of Windows and Data Volume.} Figures~\ref{fig:windows-exp} and Figure~\ref{fig:tuplenumber-exp} demonstrate that, as the number of windows or data size per window grows, the latency of \oursys increases modestly but generally stays \emph{under 10 ms}. While throughput decreases, \oursys still maintains practical performance for complex analytics. 

\hi{Number of Joins.} Figure~\ref{fig:joinnumber-exp} shows that feature scripts with more LAST JOIN operations slightly increase latency (still under 5ms) and reduce throughput marginally, remaining above 6k QPS. This is critical for feature pipelines that require joining multiple reference tables. Other methods like MySQL or DuckDB would see far larger latency spikes due to repeated index lookups and lack of incremental computation (details in Section~\ref{subsec:online}).

\vspace{-.25em}
\section{Conclusion}
\label{sec:conclusion}

In this paper, we introduced \oursys, an industrial-grade feature computation system. Unlike traditional approaches that treat online and offline feature computation separately, \oursys employed a unified query plan generator that seamlessly supports feature extraction tasks in both execution modes. \oursys introduced compilation-level optimizations, such as cycle binding and code caching. For online feature computation, \oursys incorporated advanced techniques like pre-aggregation for handling long-time windows and dynamic data adjustments to efficiently manage multi-table window unions. For offline computation, \oursys supported multi-window parallelism and time-aware data repartitioning to efficiently compute historical features at scale. \oursys proposed compact in-memory data format and stream-oriented data structures that accelerate online time-series data access. Extensive evaluation in testing and real production scenarios demonstrated \oursys outperforms existing data processing frameworks and databases in terms of execution speed and resource efficiency and maintains stable performance under varying workloads. 

\clearpage
\newpage

\bibliographystyle{ACM-Reference-Format}
\bibliography{sample-clean}

\end{document}